\documentclass[5p]{elsarticle}

\usepackage{lineno}
\usepackage{hyperref}

\usepackage{microtype}
\usepackage{color}
\usepackage{soul}
\usepackage{subcaption}
\usepackage{amssymb}
\usepackage{makeidx}  
\usepackage{caption}
\usepackage{algorithm, algpseudocode}
\usepackage{amsmath}
\usepackage{cleveref}

\biboptions{authoryear}
\algtext*{EndWhile}
\algtext*{EndIf}
\algtext*{EndFor}

\definecolor{dgreen_100}{RGB}{139, 195, 74}
\definecolor{dgreen_90}{RGB}{150, 200, 91}
\definecolor{dgreen_70}{RGB}{173, 212, 127}
\definecolor{dgreen_50}{RGB}{197, 225, 164}
\definecolor{dgreen_25}{RGB}{226, 240, 210}

\definecolor{cloud_green}{RGB}{58, 182, 78}
\definecolor{could_orange}{RGB}{255, 165, 0}

\newcommand{\inputindent}{\hspace*{\algorithmicindent}\hspace*{\algorithmicindent}\hspace{.16667em} }
\newcommand{\localvarsindent}{\inputindent\inputindent \hspace{.56em} }
\newcommand{\Returns}{\textbf{returns} }

\newcommand{\Or}{\textbf{or }}
\newcommand{\hlinew}{\rule{\linewidth}{1pt} }

\journal{Expert Systems with Applications}

\begin{document}

\begin{frontmatter}

\title{A Text Classification Framework for Simple and Effective Early Depression Detection Over Social Media Streams}

\author[unsl,conicet]{Sergio G. Burdisso\corref{cor}}
\ead{sburdisso@unsl.edu.ar}

\author[unsl]{Marcelo Errecalde}
\ead{merreca@unsl.edu.ar}

\author[inaoe]{Manuel Montes-y-G\'omez}
\ead{mmontesg@inaoep.mx}

\address[unsl]{Universidad Nacional de San Luis (UNSL), Ej\'ercito de Los Andes 950, San Luis, San Lius, C.P. 5700, Argentina}
\address[conicet]{Consejo Nacional de Investigaciones Cient\'ificas y T\'ecnicas (CONICET), Argentina}
\address[inaoe]{Instituto Nacional de Astrof\'isica, \'Optica y Electr\'onica (INAOE), Luis Enrique Erro No. 1, Sta. Ma. Tonantzintla, Puebla, C.P. 72840, Mexico}

\cortext[cor]{Corresponding author}

\begin{abstract}
With the rise of the Internet, there is a growing need to build intelligent systems that are capable of efficiently dealing with early risk detection (ERD) problems on social media, such as early depression detection, early rumor detection or identification of sexual predators.
These systems, nowadays mostly based on machine learning techniques, must be able to deal with data streams since users provide their data over time. In addition, these systems must be able to decide when the processed data is sufficient to actually classify users.
Moreover, since ERD tasks involve risky decisions by which people's lives could be affected, such systems must also be able to justify their decisions.
However, most standard and state-of-the-art supervised machine learning models are not well suited to deal with this scenario.
This is due to the fact that they either act as black boxes or do not support incremental classification\slash learning.
In this paper we introduce SS3, a novel supervised learning model for text classification that naturally supports these aspects. SS3 was designed to be used as a general framework to  deal with ERD problems.
We evaluated our model on the CLEF's eRisk2017 pilot task on early depression detection.
Most of the 30 contributions submitted to this competition used state-of-the-art methods.
Experimental results show that our classifier was able to outperform these models and standard classifiers, despite being less computationally expensive and having the ability to explain its rationale.
\end{abstract}

\begin{keyword}
Early Text Classification. Early Depression Detection. Incremental Classification. SS3. Interpretability. Explainability.
\end{keyword}

\end{frontmatter}


\section{Introduction}
Traditionally, \emph{expert systems} have been used to deal with complex problems that require the ability of human experts to be solved.
These intelligent systems usually need knowledge engineers to manually code all the facts and rules acquired from human experts through interviews, for the system's \emph{knowledge base} (KB).
Nonetheless, This manual process is very expensive and error-prone since the KB of a real expert system includes thousands of rules.
This, added to the rise of big data and cheaper GPU-powered computing hardware, are causing a major shift in the development of these intelligent systems in which machine learning is increasingly gaining more popularity.
In this context, this work introduces a machine learning framework, based on a novel white-box text classifier, for developing intelligent systems to deal with early risk detection (ERD) problems. In order to evaluate and analyze our classifier's performance, we will focus on a relevant ERD task: early depression detection.

\

Depression detection is a major public health concern. Depression is a leading cause of disability and is a major contributor to the overall global burden of disease.
Globally, the proportion of the population with depression in 2015 was estimated to be 4.4\% (more than 332 million people).
Depressive disorders are ranked as the single largest contributor to non-fatal health loss.
More than 80\% of this non-fatal disease burden occurs in low- and middle-income countries.
Furthermore, between 2005 and 2015 the total estimated number of people living with depression was increased by 18.4\% \citep{who2017}.

People with depression may experience a lack of interest and pleasure in daily activities, significant weight loss or gain, insomnia or excessive sleeping, lack of energy, inability to concentrate, feelings of worthlessness or excessive guilt and recurrent thoughts of death \citep{apa2013}.
As a matter of fact, depression can lead to suicide.
Over 800.000 suicide deaths occur every year and it is the second leading cause of death in the 15-29 years-old range; that is, every 40 s a person dies due to suicide somewhere in the world \citep{who2014}.
In richer countries, three times as many men die of suicide than women do. Globally, suicides account for 50\% of all violent deaths in men and 71\% in women \citep{who2014}.
Suicide accounted for close to 1.5\% of all deaths worldwide, bringing it into the top 20 leading causes of death in 2015 \citep{who2017}.
In the United States, as well as in other high-income countries, suicide is among the 10 leading causes of death (along with cancer, heart disease, stroke, and diabetes), additionally, from 2016 to 2017 the suicide rate increased by 3.7\% \citep{cdc2019}.

In this context, it is clear that the early risk recognition is a core component to ensure that people receive the care and social support they need. 
For many years, psychologists have used tests or carefully designed survey questions to assess different psychological constructs.
Nowadays methods for automatic depression detection (ADD) have gained increasing interest since all the information available in social media, such as Twitter and Facebook, enables novel measurement based on language use. In \citep{schwartz2015data}, it is highlighted that ``language reveals who we are: our thoughts, feelings, belief, behaviors, and personalities''. In particular, quantitative analysis of the words and concepts expressed in texts have played an important role in ADD. For instance, in \citep{de2013predicting} the written content of tweets shared by subjects diagnosed with clinical depression are analyzed and an SVM classifier is trained to predict if a tweet is depression-indicative.

A pioneering work in this area \citep{stirman2001word} used the Linguistic Inquiry and Word Count (LIWC) (an automated word counting software) and showed that it is possible to characterize depression through natural language use.
There, it is suggested that suicidal poets use more first-person pronouns (e.g., I, me, mine) and less first plural pronouns (e.g., we, ours) throughout their writing careers than non-suicidal poets.
In a similar way, depressed students are observed to use first-person singular pronouns more often, more negative emotion words and fewer positive emotion words in their essays in comparison to students who have never suffered from this disease \citep{rude2004language}.

In the context of online environments such as social media, an ADD scenario that is gaining interest, as we will see in \autoref{subsec:ADD}, is the one known as \emph{early depression detection} (EDD). In EDD the task is, given users' data stream, to detect possible depressive people \emph{as soon and accurate as possible}.

Most automatic approaches to ADD have been based on standard machine learning algorithms~\citep{guntuku2017detecting,tsugawaKKNIO15,marinelarena2017predicting}.
However, EDD poses really challenging aspects to the ``standard'' machine learning field.
The same as with any other ERD task, we can identify at least three of these key aspects: \emph{incremental classification} of sequential data, \emph{support for early classification} and, \emph{explainability}\footnote{Having the ability to explain its rationale.}.

To put the previous points in context, it is important to note that ERD is essentially a problem of \emph{analysis of sequential data}.
That is, unlike traditional supervised learning problems where learning and classification are done on ``complete'' objects, here classification (or both) must be done on ``partial'' objects which correspond to all the data sequentially read up to the present, from a (virtually infinite) data stream.
Algorithms capable of dealing with this scenario are said to support \emph{incremental learning} and/or \emph{incremental classification}.
In the present article we will focus on incremental classification since, so far, it is the only EDD scenario we have data to compare to, as we will see in \autoref{subsec:ADD}.
However, as we will see later, our approach is designed to incrementally work in both the learning and classification phases.

Classifiers supporting incremental classification of sequential data need to provide a suitable method to ``remember'' (or ``summarize'') historical information read up to the present. The informativeness level of these partial models will be critical to the effectiveness of the classifier. In addition, these models also need to provide support to a key aspect of \emph{ERD}: the decision of \emph{when} (how soon) the system should stop reading from the input stream and classify it with acceptable accuracy. This aspect, that we have previously mentioned as the \emph{supporting for early classification}, is basically a multi-objective decision problem that attempts to balance accurate and timely classifications.

Finally, \emph{explainability\slash interpretability} is another important requirement for EDD.
The same as with any other critical application in healthcare, finance, or national security, this is a domain that would be greatly benefited by models that not only make correct predictions but also facilitate understanding how those predictions are derived.
Although interpretability and explanations have a long tradition in areas of AI like \emph{expert systems} and \emph{argumentation}, they have gained renewed interest in modern applications due to the complexity and obscure nature of popular machine learning methods based on deep learning.

In the present work we propose a novel text classification model, called SS3, whose goal is to provide support for \emph{incremental classification}, \emph{early classification} and \emph{explainability} in a unified, simple and effective way.
We mainly focus on the first two aspects, measuring the SS3's effectiveness on the first publicly-available EDD task, whereas regarding the latter we present very promising results showing how SS3 is able to visually explain its rationale.

The remaining sections are organized as follows.
\autoref{sec:rel-work} presents those works that relate to ours. 
The proposed framework is introduced in \autoref{sec:framework}, firstly introducing the general idea and then the technical/formal details.
In \autoref{sec:exp} the proposed framework is compared to state-of-the-art methods used in a recent early depression detection task. \autoref{sec:analysisAndDiscussion} goes into details of the main contributions of our approach by analyzing quantitative and qualitative aspects of the proposed framework. Finally, \autoref{sec:conclusions} summarizes the main conclusions derived from this study and suggests possible future work.

\section{Related Work}
\label{sec:rel-work}

We organized the related works into 2 subsections. The first one describes works related to early classification in sequential data.
The second subsection addresses the problem of early depression detection.

\subsection{Analysis of Sequential Data: Early classification}
\label{subsec:ASD}

The analysis of sequential data is a very active research area that addresses problems where data is processed naturally as sequences or can be better modeled that way, such as sentiment analysis,  machine translation, video analytics, speech recognition, and time series processing.
A scenario that is gaining increasing interest in the classification of sequential data is the one referred to as ``early classification'', in which, the problem is to classify the data stream as early as possible without having a significant loss in terms of accuracy.

For instance, some works have addressed early text classification  by using diverse techniques like modifications of Naive Bayes  \citep{escalante2015early}, profile-based representations \citep{escalante2017early}, and Multi-Resolution Concept Representations \citep{lopez2018early}. Those approaches have focused on quantifying prediction performance of the classifiers when using partial information in documents, that is, by considering how well they behave when incremental percentages of documents are provided to the classifier.
However, those approaches do not have any mechanisms to decide \emph{when} (how soon) the partial information read is sufficient to classify the input.
Note that this is not a minor point since, for instance, in online scenarios in which users provide
their data over time, setting a manually fixed percentage of the input to be read would not be possible\footnote{If we see the input (the user's data) as a single document, this document would be virtually infinite! growing over time as the user generates new content.}.
This scenario, that we address here as the ``real'' early sequence classification problem, can be considered as a concrete multi-objective problem in which the challenge is to find a trade-off between the earliness and the accuracy of classification~\citep{xing2010brief}.

The reasons behind this requirement of ``earliness'' could be diverse. It could be necessary because the sequence length is not known in advance (e.g. online scenarios as suggested above) or, for example, if savings of some sort (e.g. computational savings) can be obtained by classifying the input in an early fashion.
However, the most important (and interesting) cases are when the delay in that decision could also have negative or risky implications.
This scenario, known as ``early risk detection'' have gained increasing interest in recent years with potential applications in rumor detection \citep{ma2015detect,ma2016detecting,kwon2017rumor}, sexual predator detection and aggressive text identification \citep{escalante2017early}, depression detection \citep{losada2017erisk, losada2016test} or terrorism detection \citep{iskandar2017terrorism}.

The key issue in real early sequence classification is that learned models usually do not provide guidance about how to decide the correct moment to stop reading a stream and classify it with reasonable accuracy.
As far as we know, the approach presented in \citep{dulac2011text} is the first to address a (sequential) text classification task as a Markov decision process (MDP) with virtually three possible actions: read (the next sentence), classify\footnote{In practice, this action is a collection of actions, one for each category \emph{c}.} and stop.
The implementation of this model relied on using Support Vector Machines (SVMs) which were trained to classify each possible action as ``good'' or ``bad'' based on the current state, \emph{s}. This state was represented by a feature vector, $\Phi(s)$, holding information about the \emph{tf-idf} representations of the current and previous sentences, and the categories assigned so far.
Although the use of MDP is very appealing from a theoretical point of view, and we will consider it for future work, the model they proposed would not be suitable for risk tasks. The use of SVMs along with $\Phi(s)$ implies that the model is a black box, not only hiding the reasons for classifying the input but also the reasons behind its decision to stop early\footnote{Since this is enforced by the reward function which in turn depends, for each state \emph{s}, on vector $\Phi(s)$.}.
The same limitations could be found in more recent works \citep{yu2017learning, yu2018fast, shen2017reasonet} also addressing the early sequence classification problem as a reinforcement learning problem but using Recurrent Neural Networks (RNNs).

Finally,~\citep{loyola2018learning} considers the decision of ``when to classify'' as a problem to be learned on its own and trains two SVMs, one to make category predictions and the other to decide when to stop reading the stream.
Nonetheless, the use of these two SVMs, again, hides the reasons behind both, the classification and the decision to stop early. Additionally, as we will see in \autoref{subsec:complexity}, when using SVM to classify a document, incrementally, the classification process becomes costly and not scalable, since the \emph{document-term matrix} has to be re-built from scratch every time new content is added.

\subsection{Early Depression Detection}
\label{subsec:ADD}

Even though multiple studies have attempted to predict or analyze depression using machine learning techniques, before \citep{losada2016test}, no one had attempted to build a public dataset in which a large chronological collection of writings, leading to this disorder, were made available to the research community.
This is mainly due to the fact that text is often extracted from social media sites, such as Twitter or Facebook, that do not allow redistribution.
On the other hand, in the machine learning community, it is well known the importance of having publicly available datasets to foster research on a particular topic, in this case, predicting depression based on language use.
That was the reason why the main goal in \citep{losada2016test} was to provide, to the best of our knowledge, the first public collection to study the relationship between depression and language usage by means of machine learning techniques. This work was important for ADD, not only for creating this publicly-available dataset for EDD experimentation but also because they proposed a measure (ERDE) that simultaneously evaluates the accuracy of the classifiers and the delay in making a prediction. 
It is worth mentioning that having a single measure combining these two aspects enabled this dataset to be used as a benchmark task in which different studies can be compared in terms of how ``early-and-accurate'' their models are.

Both tools, the dataset and the evaluation measure, were later used in the first pilot task of eRisk \citep{losada2017erisk} in which 8 different research groups submitted a total of 30 contributions.
Given that we will use this dataset for experimentation, evaluating and analyzing our results in comparison with the other 30 contributions, we will analyze them in more detail.

As observed in \citep{losada2017erisk}, among the 30 contributions submitted to the eRisk task, a wide range of different document representations and classification models were used.
Regarding document representations some research groups used simple features like standard Bag of Words \citep{trotzek2017linguistic,errecaldetemporal,fariasuach}, bigrams and trigrams \citep{errecaldetemporal,almeida2017detecting,fariasuach}, while others used more elaborated and domain-specific ones like lexicon-based features\footnote{Such as emotion words from WordNet, sentiment words from Vader, and  preexisting depression-related dictionaries.}\citep{malam2017irit,trotzek2017linguistic,sadeque2017uarizona,almeida2017detecting}, LIWIC features \citep{trotzek2017linguistic,errecaldetemporal}, Part-of-Speech tags \citep{almeida2017detecting}, statistical features\footnote{Such as the average number of posts, the average number of words per post, post timestamps, etc.}\citep{malam2017irit,almeida2017detecting,fariasuach} or even hand-crafted features \citep{trotzek2017linguistic}.
Some other groups made use of more sophisticated features such as Latent Semantic Analysis \citep{trotzek2017linguistic}, Concise Semantic Analysis \citep{errecaldetemporal}, Doc2Vec \citep{trotzek2017linguistic} or even graph-based representations \citep{villatorouam}.
Regarding classification models, some groups used standard classifiers\footnote{Such as Multinomial Naive Bayes(MNB), Logistic Regression (LOGREG), Support Vector Machine(SVM), Random Forest, Decision Trees, etc.}\citep{malam2017irit,trotzek2017linguistic,sadeque2017uarizona,errecaldetemporal,almeida2017detecting,fariasuach} while others made use of more complex methods such as different types of Recurrent Neural Networks \citep{trotzek2017linguistic,sadeque2017uarizona}, graph-based models \citep{villatorouam}, or even combinations or ensemble of different classifiers \citep{trotzek2017linguistic,sadeque2017uarizona,errecaldetemporal,almeida2017detecting}.

Another interesting aspect of this evaluation task was the wide variety of mechanisms used to decide \emph{when} to make each prediction.
Most research groups \citep{malam2017irit,trotzek2017linguistic,sadeque2017uarizona,villatorouam,errecaldetemporal,almeida2017detecting} applied a simple policy in which, the same way as in \citep{losada2016test}, a subject is classified as depressed when the classifier outputs a value greater than a fixed threshold. Some other groups~\citep{fariasuach} applied no policy at all and no early classification was performed, i.e. their classifiers made their predictions only after seeing the entire subject's history\footnote{Note that this is not a realistic approach, usually there is no such thing as a subject's ``last writing'' in real life since subjects are able to create new writings over time.}.
It is worth mentioning that some  groups \citep{malam2017irit,trotzek2017linguistic,errecaldetemporal} added extra conditions to the given policy, for instance \citep{trotzek2017linguistic} used a list of manually-crafted rules of the form: ``if output $\geq\alpha_n$ and the number of writings $\geq n$, then classify as positive'', ``if output $\leq\beta_n$ and the number of writings $\geq n$, then classify as non-depressed'', etc.

As will be highlighted and analyzed in more detail later, in \autoref{sec:analysisAndDiscussion}, none of these 30 contributions, except those based on RNN and MNB, are suitable for naturally processing data sequences since, as mentioned earlier, standard classifiers such as Logistic Regression (LOGREG), SVM,  (feedforward) Neural Network (NN), etc. are designed to work with complete and atomic document representations.
Furthermore, no contributions paid attention to the explainability of their models since all of them (even those based on RNN and MNB) act as black boxes, which we consider a key aspect when dealing with risk applications in which real people are involved.

\section{The SS3 Framework}
\label{sec:framework}

At this point, it should be clear that any attempt to address ERD problems, in a realistic fashion, should take into account 3 key requirements: \emph{incremental classification}, \emph{support for early classification}, and \emph{explainability}.
Unfortunately, to the best of our knowledge, there is no text classifier able to support these three aspects in an integrated manner.
In the remainder of this section, we will describe a new text classifier that we have created with the goal to achieve it.

Additionally, since we are introducing a new classification model, instead of going straight to the plain equations and algorithms, we have decided to include the general idea first and then, along with the equations, the ideas that led us to them.
Thus, \autoref{sec:general_operation} shows the general operation of our framework with an informative and intuitive example highlighting how the above requirements could be met. 
Finally, \autoref{sec:formal_description} goes into some technical details about how the introduced ideas are actually implemented.

\subsection{General Operation}
\label{sec:general_operation}

In this subsection, we will give an intuitive and general idea of how our framework could address the above requirements with a simple and incremental classification model that we have called ``\emph{SS3}'', which stands for Sequential S3 (Smoothness, Significance, and Sanction) for reasons that will be clear later on. Our humble approach was intended to be used as a general framework for solving the document classification problem since it is flexible enough to be instantiated in several different manners, depending on the problem.

In the rest of this subsection, we will exemplify how the SS3 framework carries out the classification and training process and how the early classification and explainability aspects are addressed. The last subsection goes into more technical details and we will study how the local and global value of a term is actually computed. As we will see, these values are the basis of the entire classification process.

\subsubsection{Classification Process}
\label{sec:classification_process}

This subsection describes how classification is carried out.
However, before we illustrate the overall process and for the sake of simplicity, we are going to assume there exist a function $gv(w,c)$ to value words in relation to categories ---and whose formal definition will be the topic of \autoref{sec:local_global_value}.
To be more specific, $gv$ takes a word $w$ and a category $c$ and outputs a number in the interval [0,1] representing the degree of $confidence$ with which $w$ is believed to \emph{exclusively} belong to $c$, for instance:

\vspace{5mm}

\begin{tabular}{@{}l@{}l} 
$gv($`$apple$'$, travel) = 0;$&$gv($`$the$'$, travel) = 0;$\\
$gv($`$apple$'$, technology) = 0.8;$&$gv($`$the$'$, technology) = 0;$\\
$gv($`$apple$'$, business) = 0.4;$&$gv($`$the$'$, business) = 0;$\\
$gv($`$apple$'$, food) = 0.75;$  & $gv($`$the$'$, food) = 0;$\\
\end{tabular}

\vspace{5mm}

Where $gv(w, c) = v$ is read as ``$w$ has a \emph{global value} of $v$ in $c$'' or, alternatively, ``the \emph{global value} of $w$ in $c$ is $v$''.
For example, $gv($`$apple$'$, technology) = 0.8$ is read as ``\emph{apple} has a \emph{global value} of 0.8 in \emph{technology}''.
Additionally, we will define $gv(w)=(gv(w,c_0), gv(w,c_1), \dots, gv(w,c_k))$ where $c_i \in C$, and $C$ denotes the set of all the categories.
That is, when $gv$ is only applied to a word it outputs a vector in which each component is the \emph{global value} of that word for each category $c_i$.
For instance, following the above example, we have:

\vspace{5mm}

\begin{tabular}{@{}l p{0.5\linewidth}}
    $gv(apple) = (0, 0.8, 0.4, 0.75);$ & $gv(the) = (0, 0, 0, 0);$
\end{tabular}

\vspace{5mm}

The vector $gv(w) = \overrightarrow{v}$ will be called ``\emph{confidence vector} of $w$''; thus $(0, 0, 0, 0)$ is the \emph{confidence vector} of the word ``the'' in the example above. Note that each category $c_i$ is assigned to a fixed position $i$ in the output vector ---in this example, the first position corresponds to $travel$, the second to $technology$, and so on.

\begin{figure*}[t!] 
    \centering
    \includegraphics[width=190mm]{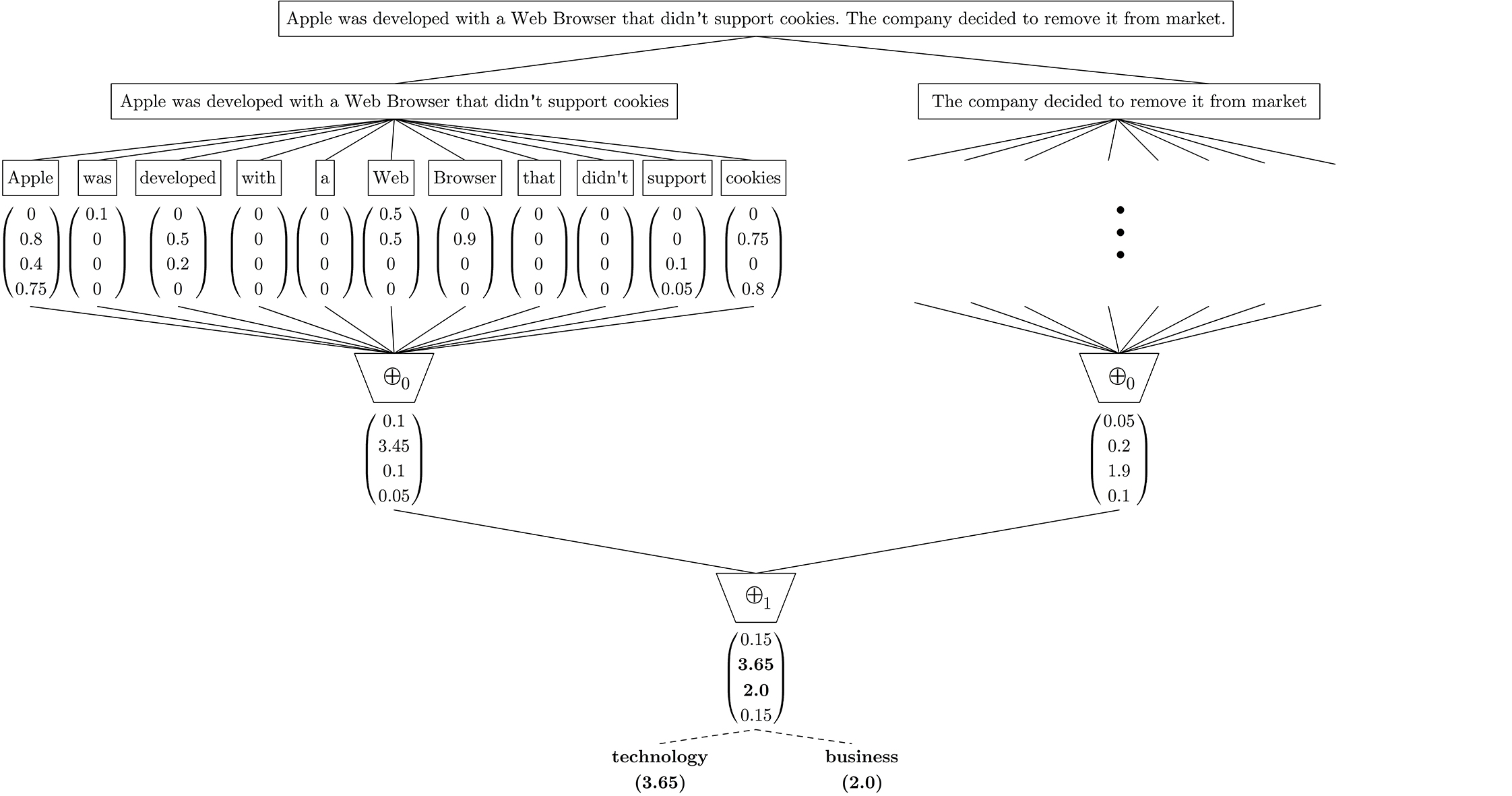}
    \caption{Classification process for a hypothetical example document ``Apple was developed with a Web Browser that didn't support cookies. The company decided to remove it from the market''.
In the first stage, this document is split into two sentences (for instance, by using the dot as a delimiter) and then each sentence is also split into single words.
In the second stage,  \emph{global values} are computed for every word to generate the first set of \emph{confidence vectors}.
Then all of these word vectors are reduced by the $\oplus_0$ operator to sentence vectors, $(0.1, 3.45, 0.1, 0.05)$ and $(0.05, 0.2, 1.9, 0.1)$ for the first and second sentence respectively.
After that, these two sentence vectors are also reduced by another operator ($\oplus_1$, which in this case is the addition operator) to a single \emph{confidence vector} for the entire document, $(0.15, 3.65, 2.0, 0.15)$.
Finally, a policy is applied to this vector to make the classification ---which in this example was to select \emph{technology}, the category with the highest value, and also \emph{business} because its value was ``close enough'' to \emph{technology}'s.}
    \label{fig:mapreduce}
\end{figure*}

\begin{figure}[t!] 
    \centering
    \includegraphics[width=90mm]{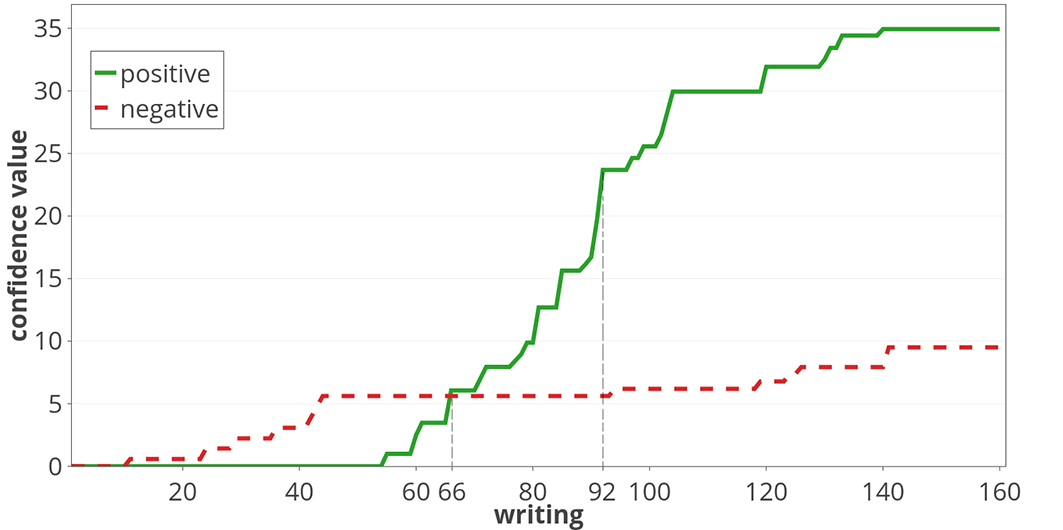}
    \caption{subject 9579's positive and negative confidence value variation over time. Time is measured in writings and it could be further expanded as more writings are created by the subject over time.}
    \label{fig:subject9579_writings}
\end{figure}

Now that the needed basic definitions and terminology have been introduced, we are ready to describe the overall classification process, which is illustrated with an example in \autoref{fig:mapreduce}.
Classification can be thought of as a 2-phase process.
The first phase starts out by splitting the given input (usually a single document) into multiple blocks, then each block is in turn repeatedly divided into smaller units until words are reached.
At the end of this phase, we have converted the previously ``flat'' input into a hierarchy of blocks.
In practice, a document will be typically divided into paragraphs, paragraphs into sentences and sentences into words.
Additionally, we will say that words are at level 0 in this hierarchy, sentences at level 1, paragraphs at level 2, and so on.
In the second phase, the $gv$ function is applied to each word to obtain the level 0 \emph{confidence vectors}, which then are reduced by means of a \emph{summary operator} to generate the next level's \emph{confidence vectors}.
This reduction process is recursively propagated up to higher-level blocks until a single \emph{confidence vector} is generated for the whole input.
Finally, the actual classification is performed based on the values of this single \emph{confidence vector} ---some policy must be used, e.g. the category with the maximum value.
Note that in the example shown in \autoref{fig:mapreduce}, \emph{summary operators} are denoted by $\oplus_j$, where $j$ denotes the level, to highlight the fact that each level (e.g. words, sentences, etc.) could have a different \emph{summary operator}
---for instance, $\oplus_0$ could be addition, $\oplus_1$ maximum (i.e. max pooling), $\oplus_2$ average (i.e. mean pooling), etc. Moreover, any function of the form $f:2^{\mathbb{R}^n}\mapsto\mathbb{R}^n$ could be used as a \emph{summary operator}.

\begin{algorithm*}
\small
\caption{\small General multi-label classification algorithm.
$MAX\_LEVEL$ is a constant storing the maximum hierarchy level when partitioning the document. For instance, it should be 3 when working with the paragraph-sentence-and-word partition.
\textproc{Global-Value} is the $gv$ function.
\textproc{Map} applies \textproc{Classify-At-Level}($block$, $n-1$) to every $block$ in $blocks$ and returns a list of resultant vectors.
\textproc{Reduce} reduces $blocks\_cvs$ to a single vector by applying the $\oplus_{n-1}$ operator cumulatively to the vectors in $blocks\_cvs$.}
\label{alg:classification}
\begin{algorithmic}[h]
\Statex
\Function{Classify}{$text$} \Returns a set of category indexes
	\State \textbf{input:} $text$, the sequence of one or more symbols
	\State \textbf{local variables:} $\overrightarrow{c}$, the document \emph{confidence vector}
	\State
	\State $\overrightarrow{c} \gets$ \Call{Classify-At-Level}{$text$, $MAX\_LEVEL$}
	\State \Return a set of indexes selected by applying a policy, $\pi$, to $\overrightarrow{c}$ 
\EndFunction
\Statex \hrulefill
\Function{Classify-At-Level}{$text$, $n$} \Returns a confidence vector
	\State \textbf{input:} $text$, a sequence of symbols
	\State \textbf{local variables:} $blocks$, a list of smaller blocks of the text
	\State \localvarsindent $blocks\_cvs$, block \emph{confidence vectors} list
	\State
	\If{$n$ == 0} \Comment i.e. if $text$ is equal to a single symbol
		\State \Return \Call{Global-Value}{$text$} 
	\Else
		\State $blocks \gets$  split $text$ into smaller  units based on a level $n$ delimiter
		\State $blocks\_cvs \gets$ \Call{Map}{}(\Call{Classify-At-Level}{}, $blocks$, $n - 1$)
		\State \Return \Call{Reduce}{}(\Call{$\oplus_{n-1}$}{}, $blocks\_cvs$)
	\EndIf
\EndFunction
\end{algorithmic}
\end{algorithm*}

It is worth mentioning that with this simple mechanism it would be fairly straightforward to justify when needed, the reasons of the classification by using the values of \emph{confidence vectors} in the hierarchy, as will be illustrated with a visual example at the end of \autoref{sec:analysisAndDiscussion}.
Additionally, the classification is also incremental as long as the \emph{summary operator} for the highest level can be computed in an incremental fashion
---which is the case for most common aggregation operations such as addition, multiplication, maximum or even average\footnote{In case of average it would be necessary to store, in addition to a vector with the sum of all previous \emph{confidence vectors}, their number.}.
For instance, suppose that later on, a new sentence is appended to the example shown in \autoref{fig:mapreduce}.
Since $\oplus_1$ is the addition, instead of processing the whole document again, we could update the already computed vector, $(0.15, 3.65, 2.0, 0.15)$, by adding it to the new sentence \emph{confidence vector}---
Note that this incremental classification, in which only the new sentence needs to be processed, would produce exactly the same result as if the process were applied to the whole document again each time.

Another important aspect of this incremental approach is that since this \emph{confidence vector} is a value that ``summarizes the past history'', keeping track of \emph{how} this vector changes over time should allow us to derive simple and clear rules to decide \emph{when} the system should make an early classification. 
As an example of this, suppose we need to classify a social media user (i.e. a subject) as depressed (positive) or non-depressed (negative) based on his/her writings.
Let us assume that this user is the subject 9579, he/she is depressed, and that the change of each \emph{confidence vector} component over time (measured in writings) is the one shown in \autoref{fig:subject9579_writings}.
We could make use of this ``dynamic information'' to apply certain policies to decide when to classify subjects as depressed.
For example, one of such a policy would be ``classify a subject as positive when the accumulated positive value becomes greater than the negative one'' ---in which case, note that our subject would be classified as depressed after reading his/her 66th writing.
Another (more elaborated) policy could have taken into account how fast the positive value grows (the slope) in relation with the negative one, and if a given threshold was exceeded, classify subjects as depressed ---in such case our subject could have been classified as depressed, for instance, after reading his/her 92nd writing.
Note that we could also combine multiple policies as we will see in \autoref{sec:analysisAndDiscussion}.

\subsubsection{Training Process}
\label{sec:training_process}

This brief subsection describes the training process, which is trivial. Only a dictionary of term-frequency pairs is needed for each category.
Then, during training, dictionaries are updated as new documents are processed ---i.e. unseen terms are added and frequencies of already seen terms are updated.

\vspace{5mm}

Note that with this simple training method there is no need neither to store all documents nor to re-train from scratch every time a new training document is added, making the training incremental\footnote{Even new categories could be dynamically added.}. Additionally, there is no need to compute the document-term matrix because, during classification, $gv$ can be dynamically computed based on the frequencies stored in the dictionaries ---although, in case we are working in an offline fashion and to speed up classification, it is still possible to create the document-term matrix holding the $gv$ value for each term.
Finally, also note that training computation is very cheap since involves only updating term frequencies i.e only one addition operation is needed.

\subsection{Formal/technical Description}
\label{sec:formal_description}

This section presents more formally the general and intuitive description given in the previous section.

\subsubsection{Classification and Training} In \autoref{alg:classification} is shown the general multi-label classification algorithm which carries out the process illustrated earlier in \autoref{sec:classification_process}.
Note that this algorithm can be massively parallelized since it naturally follows the Big Data programming model \emph{MapReduce}~\citep{dean2008}, giving the framework the capability of effectively processing very large volumes of data. 
In \autoref{alg:training} is shown the training process described earlier. Note that the line calling the \textproc{Update-Global-Values} function, which calculates and updates all global values, is only needed if we want to construct the document-term matrix to work in the standard batch-like way.
Otherwise, it can be omitted since, during classification, $gv$ can be dynamically computed based on the frequencies stored in the dictionaries.
It is worth mentioning that this algorithm could be easily parallelized by following the MapReduce model as well ---for instance, all training documents could be split into batches, then frequencies locally calculated within each batch, and finally, all these local frequencies summed up to obtain the total frequencies.

\begin{algorithm*}
\small
\caption{\small Learning Algorithm.}
\label{alg:training}
\begin{algorithmic}[h]
\Statex
\Procedure{Learn-From-Dataset}{$dataset$}
	\State \textbf{Input:} $dataset$, a list of labeled documents
	\State
	\For{\textbf{each} $document$ \textbf{in} $dataset$}
		\State \Call{Learn-New-Document}{$document$.TEXT, $document$.CATEGORY}
	\EndFor
	\State \Call{Update-Global-Values}{}() \Comment{this line is optional}
\EndProcedure
\Statex \hrulefill
\Procedure{Learn-New-Document}{$text$, $category$}
	\State \textbf{input:} $text$, the sequence of words in the document
	\State \inputindent $category$, the category the document belongs to
	\State
	\For{\textbf{each} $word$ \textbf{in} $text$}
		\If{$word$ $\notin$ $category$.DICTIONARY}
			\State add $word$ to $category$.DICTIONARY
		\EndIf
		\State $category$.DICTIONARY[$word$] $\gets$ $category$.DICTIONARY[$word$]$ + 1$
	\EndFor
\EndProcedure
\end{algorithmic}
\end{algorithm*}

\subsubsection{Local and Global Value of a Word}
\label{sec:local_global_value}

Our approach to calculating $gv$, as we will see later, tries to overcome some problems arising from the valuation of words only based on local information to a category. This is carried out by, firstly, computing a word \emph{local value} ($lv$) for every category, and secondly, combining them to obtain the \emph{global value} of the word in relation to all the categories.

More precisely, the \emph{local value} should be a function such that $lv(w, c) \propto P(w|c)$ i.e. the \emph{local value} of $w$ in $c$ should be proportional to the probability of $w$ occurring, given the category $c$. Therefore, $lv$ will be defined by:

\begin{equation}
\label{eq:lv-first}
lv(w, c) = \frac{P(w|c)}{P(w_{max}|c)}
\end{equation}

Instead of simply having $lv(w, c) = P(w|c)$, we have chosen to divide it by the probability of the most frequent word in $c$.
This produces two positive effects:
(a) $lv$ is normalized and the most probable word will have a value of 1, and more importantly,
(b) words are now valued in relation to how close they are to the most probable one.
Therefore, no matter the category, all \emph{stop words} (such as ``the``, ``of``, ``or'', etc.) will always have a value very close, or equal, to 1.

Note that this allows us to compare words across different categories since their values are all normalized in relation to stop words, which should have a similar frequency across all the categories\footnote{Note that we are assuming here that we are working with textual information in which there exist highly frequent elements that naturally have similar frequency across all categories (e.g. such as stop words).}.
However, our current definition of $lv$ implicitly assumes that the proportionality $lv(w, c) \propto P(w|c)$ is direct, which is not always true so we will define $lv$ more generally as follows:

$$lv_\sigma(w, c) = \bigg(\frac{P(w|c)}{P(w_{max}|c)}\bigg)^\sigma$$

\vspace{5mm}

Which, after estimating the probability, $P$, by an analytical \emph{Maximum Likelihood Estimation}(MLE) derivation, leads us to the actual definition:

\vspace{5mm}

\begin{equation}
lv_\sigma(w, c) = \bigg(\frac{tf_{w,c}}{max\{tf_c\}}\bigg)^\sigma
\end{equation}

\vspace{5mm}

Where $tf_{w,c}$ denotes the frequency of $w$ in $c$ and $max\{tf_c\}$ the maximum frequency seen in $c$. The value $\sigma  \in (0, 1]$ is the first hyper-parameter of our model, called ``smoothness'', and whose role is twofold:
\begin{itemize}
    \item Control how fast grows the \emph{local value} of a word in relation to how close it is to the most probable one; e.g. when $\sigma=1$, $lv$ grows linearly proportional to $P(w|c)$.
    \item Control the smoothness of the distribution of words which otherwise, by the empirical Zipf's law \citep{zipf1949human, powers1998applications}, will have a very small group of highly frequent words overshadowing important ones.
\end{itemize}

\begin{figure}[t]
    \begin{subfigure}{0.5\textwidth}
        \centering
        \includegraphics[width=90mm]{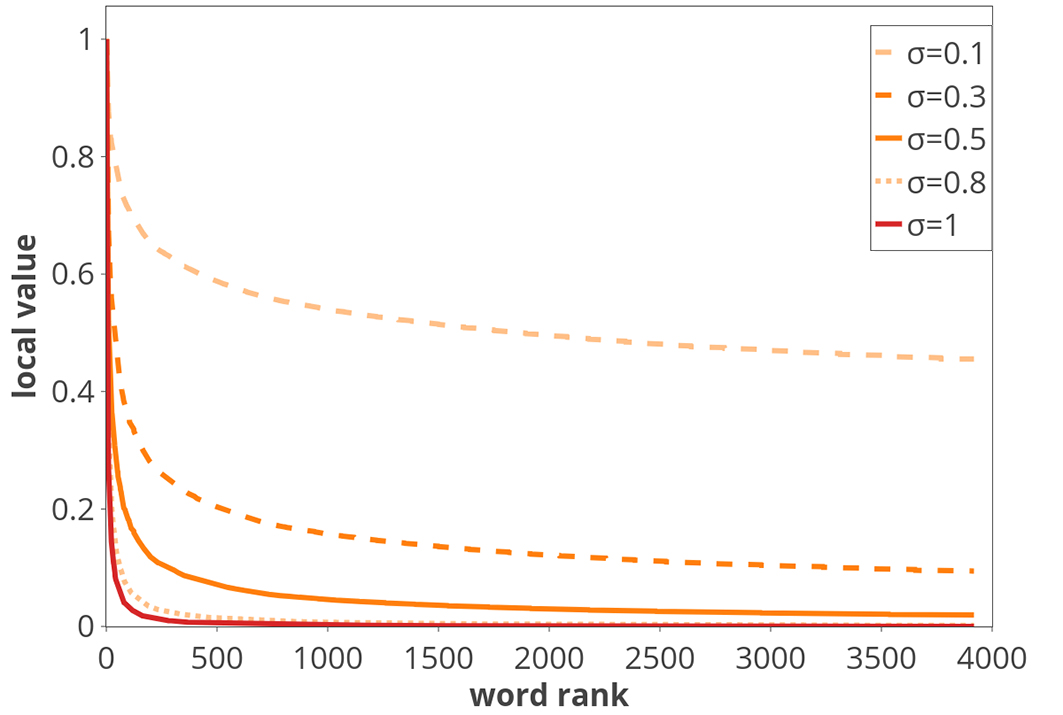}
    \end{subfigure}
\caption{
    word-\emph{local value} diagram for 5 different values of $\sigma$: 1, 0.8, 0.5, 0.3 and 0.1.
    The abscissa represents individual words arranged in order of frequency.
    Note that when $\sigma=1$, $lv_1$ (red line) matches the shape of the raw frequency (the actual word distribution), however, as $\sigma$ decreases, the curve becomes smoother; reducing the gap between the highest and the lowest values.
}
\label{fig:sigma}
\end{figure}

These two items are illustrated with an example in \autoref{fig:sigma} ---a good value for $\sigma$ should be around 0.5, which would be approximately equivalent to taking the square root of the \autoref{eq:lv-first}.

\vspace{5mm}

Now that we are able to compute word \emph{local values}, we are going to define its \emph{global value} based on them, as follows:

\begin{equation}
gv(w, c) = lv_\sigma(w, c)\cdot sg_{\lambda}(w, c)\cdot sn_\rho(w, c)    
\end{equation}

Where $sg$ and $sn$ are functions of the form $f:W\times C\mapsto [0, 1]$.
As we will see, the former decreases $lv$ in relation to the global significance of $w$, and the latter sanctions it, in relation to the number of categories for which $w$ is significant.
Additionally, the values  $\lambda, \rho \in \mathbb{R}^+$ referred to as ``significance'' and ``sanction'' respectively, are the other two hyper-parameters of our model.

\vspace{5mm}

In order to represent the significance of a word, $w$, with respect to a category, $c$, $sg(w, c)$ should be a function such that:
(a) it outputs a value close to 1 when $lv(w, c)$ is significantly greater than $lv(w, c_i)$, for most other categories $c_i$; and
(b) it outputs a value close to 0 when all $lv(w, c_i)$ are close to each other, for all $c_i$.
For instance, $lv($`$the$'$, c_i)$ probably will be a similarly large value for all categories $c_i$, whereas $lv($`$bread$'$, food)$ probably will be greater than most  $lv($`$bread$'$, c_i)$, for other categories; hence $sg($`$the$'$, c_i)$ should be close to 0 and $sg($`$bread$'$, food)$ close to 1.
In general, we could model this behavior by using any sigmoid function, as follows:

$$sg_{\lambda}(w, c) = sigmoid\Big(lv(w, c) - \widetilde{LV}_w, \ \lambda\cdot MAD_w\Big)$$

Such that:
\begin{enumerate}
    \item $sigmoid(d, l)\approx 1$ if $d\geq l$; and
    \item $sigmoid(d, l)\approx 0$ if $d\leq0$.
\end{enumerate}

Where $LV_w = \{lv(w, c_i) | c_i\in C\}$ i.e. the set of all \emph{local values} of $w$;
$\widetilde{LV}_w$ denotes the median of $LV$;
$MAD_w = median(|lv(w,c_i)-\widetilde{LV}_w|)$ i.e. the \emph{Median Absolute Deviation} of $LV_w$.
Additionally, note that the hyper-parameter $\lambda$\footnote{
    if $\lambda$ is approximately close to 1.4826, then $\lambda\cdot MAD_w$ is approximately equal to the \emph{standard deviation} of $LV_w$, thus perhaps setting $\lambda \approx 3\times1.4826$ would be a good value, as long as $LV_w$ has a normal distribution
}
controls how far the \emph{local value} must deviate from the median to be considered significant i.e. the closer $lv(w, c)-\widetilde{LV}_w$ to $\lambda\cdot MAD_w$, the closer the $sigmoid$ to 1, and therefore, also the closer $sg_{\lambda}(w, c)$ to 1 ---which is the desired behavior.

In particular, we have decided to use $tanh$ as the $sigmoid$ function, hence $sg$ is defined by:

\begin{equation}
    sg_{\lambda}(w, c) = \frac{1}{2}tanh\Big(4\frac{(lv(w, c) - \widetilde{LV}_w)}{\lambda\cdot MAD_w} - 2\Big) + \frac{1}{2}
\end{equation}

\vspace{5mm}

Finally, we need to define $sn$, the sanction function, which will proportionally decrease the \emph{global value} of $w$, in relation to the number of categories for which $w$ is significant. Hence $sn$ should be a function such that: (a) when $w$ is significant (i.e. $sg_{\lambda}(w, c) \approx 1$) to only one category $c$, $sn(w, c)$ should be equal to 1; (b) the greater the  number of categories $w$ is significant to, the lower the value of $sn(w, c)$. Therefore, we have defined $sn$ by:

\begin{equation}
\label{eq:san}
    sn_\rho(w, c) = \left(\frac
        {|C| - \left(\hat{C}_{wc} + 1\right)}
        {\left(|C|-1\right)\left(\hat{C}_{wc} + 1\right)}
    \right)^\rho
\end{equation}

Where $|C|$ denotes the number of categories and,
$$\hat{C}_{wc} = \sum_{c_i\in C-\{c\}}sg_{\lambda}(w, c_i)$$

i.e. $\hat{C}_{wc}$ is equal to the summation of $sg_{\lambda}(w, c_i)$ for all categories in $C$ except $c$.
Note that, for instance, when extreme cases are met, \autoref{eq:san} behaves properly;
namely, when $w$ is significant to almost all categories, $\hat{C}_{wc}\approx |C|-1$, and thus $sn(w, c)\approx 0$;
and when $w$ is significant to only one category, $c$, $\hat{C}_{wc} = 0$, and thus $sn(w, c)= 1$.

The hyper-parameter $\rho$\footnote{
    setting $\rho\approx 1$ probably is a good starting point, although we can adjust this value in relation to how overlapped categories are.
} controls how severe the sanction is, in proportion to the number of significant categories.  

\vspace{5mm}

To conclude this section, let us introduce a simple example to illustrate how the \emph{global value} is a percentage of its \emph{local value} given by its significance ($sg$) and sanction ($sn$). Suppose we have the following three categories $C = \{f, t, b\}$ for $food$, $tech$, and $business$ respectively, then:

\begin{table*}[t]
\centering
\caption{Summary of the task data}
\label{tab:dataset}
\begin{tabular}{l c c|c c}
\hline
& \multicolumn{2}{c}{Train} & \multicolumn{2}{c}{Test}\\
& Depressed & Control & Depressed & Control \\ \hline
No. of subjects & 83 & 403 & 52 & 349 \\
No. of submissions & 30,851 & 264,172 & 18,706 & 217,665 \\
Avg. No. of submissions per subject & 371.7 & 655.5 & 359.7 & 623.7 \\
Avg. No. of days from first to last submission & 572.7 & 626.6 & 608.3 & 623.2 \\
Avg. No. of words per submission & 27.6 & 21.3 & 26.9 & 22.5 \\ \hline

\hline
\end{tabular}
\end{table*}

\begin{itemize}
    \item {
        For stopwords, like `the', we would have, regardless of the category $c\in C$, something like:

        \vspace{5mm}

        \begin{tabular}{l@{}l}
            $gv($`$the$'$, c)$  & $= lv($`$the$'$, c)\times sg($`$the$'$, c)\times sn($`$the$'$, c)$\\
                            & $= lv($`$the$'$, c)\times 0.05\times 1$\\
                            & $= lv($`$the$'$, c)\times 0.05$\\
                            & $= 0.92\times 0.05 = 0.04$
        \end{tabular}

        \vspace{5mm}
        
        While the \emph{local value} of `the' is 0.92, its final \emph{global value} turned out to be 0.04 (5\% of its \emph{local value}).
        This is due to the fact that $lv($`$the$'$, c)$ is similarly high for all categories, and by definition, the significance function should be close to 0 ---in this case, $sg($`$the$'$, c)= 0.05$.
    }
    \vspace{5mm}
    \item {
        For a word that is mainly significant to a single category, in this case, `bread' to food, we would have something like:
        
        \vspace{5mm}
        
        \begin{tabular}{l@{}l}
            $gv($`$bread$'$, f)$ & $= lv($`$bread$'$, f)\times sg($`$bread$'$, f)$\\ 
            					& \ \ \ $\times sn($`$bread$'$, f)$\\
                                & $= lv($`$bread$'$, f)\times 0.99\times 0.95$\\
                                & $= lv($`$bread$'$, f)\times 0.94$\\
                                & $= 0.65\times 0.94 = 0.61$
        \end{tabular}
        
        \vspace{5mm}
        
        The \emph{global value} is almost identical to its \emph{local value} (about 94\%). This is due to the word being significant ($sg\approx 1$) \emph{only} to food ($sn\approx 1$).
    }
    \vspace{5mm}
    \item {
        For a word that is significant to more than a single category, we would have something like:
        
        \vspace{5mm}
        
        \begin{tabular}{l@{}l}
            $gv($`$apple$'$, t)$ & $= lv($`$apple$'$, t)\times sg($`$apple$'$, t)$\\
            					& \ \ \ $\times sn($`$apple$'$, t)$\\
                                & $= lv($`$apple$'$, t)\times 0.85\times 0.6$\\
                                & $= lv($`$apple$'$, t)\times 0.51$\\
                                & $= 0.7\times 0.51 = 0.35$
        \end{tabular}
        
        \vspace{5mm}
        
        In this case, the \emph{global value} ended up being about 51\% of its \emph{local value}.
        Note that while `apple' is quite significant to $tech$ ($sg=0.85$), it must also be significant to some of the other categories, at least to a certain degree, because it is being moderately sanctioned ($sn=0.6$).
    }
\end{itemize}

It is interesting to notice that Multinomial Naive Bayes can be seen as one possible instance of the SS3 framework. Namely, when $gv(w,c) = lv(w,c) = log\ P(w|c)$ and $\oplus_j = addition$, for all $j$.
However, this instance of SS3 would not effectively fulfill our goals since terms would be valued simply and solely by their local raw frequency, which is precisely the problem that $gv$ computation tries to overcome.
For instance, under this ``local view'' of words, highly discriminatory words for a particular class would be overshadowed by any more frequent word in the class (e.g. frequent words common to all classes).
This disfavors both the power to describe what words helped to make the decision and usually the performance as well---other types of models, such as SVM, frequently outperform MNB.

\section{Experimental Evaluation}
\label{sec:exp}

In this section, we cover the experimental analysis of SS3, the proposed approach.
The next subsection briefly describes the pilot task and the dataset used to train and test the classifiers.
In \autoref{sec:eval_metric} we will introduce the time-aware metric used to evaluate the effectiveness of the classifiers, in relation to the time taken to make the decision. Finally, \autoref{sec:exp_and_results} describes the different types of experiments carried out and the obtained results.

\subsection{Dataset and Pilot Task}
\label{sec:data_sets}

Experiments were conducted on the CLEF 2017\footnote{\url{http://clef2017.clef-initiative.eu}} eRisk pilot task\footnote{\url{http://early.irlab.org/2017/task.html}}, on \emph{early risk detection of depression}.
This pilot task focused on sequentially processing the content posted by users on Reddit\footnote{\url{https://www.reddit.com}}.
The dataset used in this task, which was initially introduced and described in \citep{losada2016test}, is a collection of writings (submissions) posted by users; here users will also be referred to as ``subjects''.
There are two categories of subjects in the dataset, \emph{depressed} and \emph{control} (non-depressed).
Additionally, in order to compare the results among the different participants, the entire dataset was split into two sets: a training set and a test set.
The details of the dataset are presented in \autoref{tab:dataset}. Note that the dataset is highly unbalanced, namely, only 17\% of the subjects in the training set are labeled as \emph{depressed}, and 12.9\% in the test set.
 
It is important to note that, as it is described in Section 2.2 of \citep{losada2016test}, to construct the depression group, authors first collected users by doing specific searches on Reddit (e.g. ``I was diagnosed with depression'') to obtain self-expressions of depression diagnoses, and then they manually reviewed the matched posts to verify that they were really genuine.
According to the authors, this manual review was strict, expressions like ``I have depression'', ``I think I have depression'', or ``I am depressed'' did not qualify as explicit expressions of a diagnosis. They only included a user into the depression group when there was a clear and explicit mention of a diagnosis (e.g., ``In 2013, I was diagnosed with depression'', ``After struggling with depression for many years, yesterday I was diagnosed''). That introduces the possibility of having some noise in both categories of the collected data, therefore, from now on, when we refer to ``depressed''  it should be interpreted as ``possibly diagnosed with depression''.

In this pilot task, classifiers must decide, as \emph{early} as possible, whether each user is depressed or not based on his/her writings.
In order to accomplish this, during the test stage and in accordance with the pilot task definition, the subject's writings were divided into 10 \emph{chunks} ---thus each \emph{chunk} contained 10\% of the user's history. Then, classifiers were given the user's history, one chunk at a time, and after each chunk submission, the classifiers were asked to decide whether the subject was depressed, not depressed or that more chunks need to be read.

\subsection{Evaluation Metric}
\label{sec:eval_metric}

Standard classification measures such as $F_1$-measure ($F_1$), Precision ($\pi$) and Recall ($\rho$) are time-unaware.
For that reason, in the pilot task, the measure proposed in \citep{losada2016test} was also used, called \emph{Early Risk Detection Error} (ERDE) measure, which is defined by:

$$
  ERDE_o(d,k) = \left\{
     \begin{array}{@{}l@{\thinspace}l}
        c_{fp} &\ if\ d=p\ AND\ truth=n\\
        c_{fn} &\ if\ d=n\ AND\ truth=p\\
        lc_o(k)\cdot c_{tp} &\ if\ d=p\ AND\ truth=p\\
        0 &\ if\ d=n\ AND\ truth=n\\
     \end{array}
   \right.
$$
Where the sigmoid \emph{latency cost function}, $lc_o(k)$ is defined by:
$$lc_o(k) = 1 - \frac{1}{1+e^{k - o}}$$

The delay is measured by counting the number ($k$) of distinct textual items seen before making the binary decision ($d$) which could be positive ($p$) or negative ($n$).
The $o$ parameter serves as the ``deadline'' for decision making, i.e. if a correct positive decision is made in time $k > o$, it will be taken by $ERDE_o$ as if it were incorrect (false positive).
Additionally, in the pilot task, it was also set $c_{fn} = c_{tp} = 1$ and $c_{fp}=\frac{52}{401}=0.129$. Note that $c_{fp}$ was calculated by the number of depressed subjects divided by the total subjects in the test set.

\subsection{Implementation details}
\label{sec:implementation}

SS3 was manually coded in \emph{Python 2.7} using only built-in functions and data structures, e.g. a \emph{dict} to store the category's dictionary or \emph{map} and \emph{reduce} functions to (locally) simulate a \emph{MapReduce} pattern.\footnote{We release our implementation as a Python package called PySS3~\citep{burdisso2019pyss3}.} 
Since this paper focuses on early detection, not computing nor large-scale classification, we did not perform a real \emph{MapReduce} implementation.
Moreover, since in \autoref{subsec:exp-scenario2} we are also reporting the computation time taken by all the other classifiers and all of them must share the same type of implementation, implementing a \emph{MapReduce} version would not have been fair.
Thus, all these other models were also implemented in \emph{Python 2.7}, using the \emph{sklearn} library\footnote{\href{https://scikit-learn.org/}{https://scikit-learn.org/}}, version 0.17. Vectorization was done with the \emph{TfidfVectorizer} class, with the standard English stop words list. Additionally, terms having a document frequency lower than 20 were ignored. Finally, classifiers were coded using their corresponding \emph{sklearn} built-in classes, e.g. \emph{LogisticRegression, KNeighborsClassifier, MultinomialNB}, etc.

\subsection{Experiments and Results}
\label{sec:exp_and_results}
This subsection describes the experimental work, which was divided into two different scenarios.
In the first one, we performed experiments in accordance with the original eRisk pilot task definition, using the described \emph{chunks}.
However, since this definition assumes, by using chunks, that the total number of user's writings is known in advance\footnote{Which is not true when working with a dynamic environment, such as Social Media.}, we decided to also consider a second type of experiment, simulating a more realistic scenario, in which user's history was processed as a stream, one writing at a time.

\subsubsection{Scenario 1 - original setting, incremental chunk-by-chunk classification}
Since there were only two, barely overlapped, categories, we decided to start by fixing the SS3 framework's $\lambda$ and $\rho$ hyper-parameters to 1. In fact, we also carried out some tests with other values that improved the \emph{precision} (or \emph{recall}) but worsened the ERDE measure.
Model selection was done by 4-fold cross-validation on the training data minimizing the ERDE$_{50}$ measure while applying a \emph{grid search} on the $\sigma$ hyper-parameter. 
This \emph{grid search} was carried out at three different levels of precision. In the first level, $\sigma$ took values from $0.5\pm k\cdot10^{-1}$, with $k\in [0, 5]$. Once the best value of $\sigma$ was found, let us say $\widehat{\sigma_1}$, we started a second-level \emph{grid search} in which $\sigma$ took values from $\widehat{\sigma_1}\pm k\cdot10^{-2}$. Finally, a third search was applied around the new best value, $\widehat{\sigma_2}$, where $\sigma$ was set to $\widehat{\sigma_2}\pm k\cdot10^{-3}$, also with $k\in [0, 5]$. 

\begin{table}[t]
\centering
\caption{Results on the test set in accordance with the original eRisk pilot task (using chunks).}
\label{tab:erisk-chunks}
\begin{tabular}{l@{}c || l@{}c}
\hline
& $ERDE_5\blacktriangledown$ & & $ERDE_{50}\blacktriangledown$\\ \hline
NLPISA & 15.59\% & NLPISA & 15.59\%\\
CHEPEA & 14.75\% & LyRE & 13.74\%\\
GPLC & 14.06\% & CHEPEA & 12.26\%\\
LyRE & 13.74\% & GPLC & 12.14\%\\
UNSLA & 13.66\% & UQAMD & 11.98\%\\
UQAMD & 13.23\% & UArizonaD & 10.23\%\\
UArizonaB & 13.07\% & FHDO-BCSGA & 9.69\%\\
FHDO-BCSGB & 12.70\% & UNSLA & 9.68\%\\
SS3$^\Delta$ & 12.70\% & SS3 & 8.12\%\\
SS3 & \textbf{12.60\%} & SS3$^\Delta$ & \textbf{7.72\%}\\ \hline

\hline
\end{tabular}
\end{table}

\begin{table*}[t!]
\centering
\caption{Results on the test set using a more realistic scenario in which writings are processed sequentially.}
\label{tab:erisk-writings}
\begin{tabular}{l@{}c c c c c c c c c c}
\hline
& $ERDE_5$ & $ERDE_{10}$ & $ERDE_{30}$ & $ERDE_{50}$ & $ERDE_{75}$ & $ERDE_{100}$ & $F_1$ & $\pi$ & $\rho$ & Time \\ \hline

LOGREG & 11.7\% & 10.9\% & 9.4\% & 7.5\% & 6.3\% & 5.8\% & 0.53 & 0.41 & 0.75  & 71.3m \\
SVM & 12.0\% & 10.9\% & 9.1\% & \textbf{7.2}\% & 6.1\% & 6.0\% & \textbf{0.55} & \textbf{0.47} & 0.69 & 73.9m \\
MNB & \textbf{10.6\%} & 10.4\% & 10.4\% & 10.4\% & 10.1\% & 10.1\% & 0.24 & 0.14 & \textbf{1} & 17.5m \\
KNN & 12.6\% & 10.4\% & 8.5\% & 8.2\% & 7.9\% & 7.7\% & 0.35 & 0.22 & 0.90 & 100.6m \\
SS3 & 11.0\% & \textbf{9.8\%} & \textbf{8.0\%} & \textbf{7.2\%} & \textbf{5.8\%} & \textbf{5.5\%} & 0.54 & 0.42 & 0.77 & \textbf{3.7m} \\
SS3$^\Delta$ & 11.1\% & 9.9\% & 8.1\% & 7.3\% & 5.9\% & 5.6\% & \textbf{0.55} & 0.42 & 0.81 & \textbf{3.7m} \\ \hline

\hline
\end{tabular}
\end{table*}

After the \emph{grid search}, using the hyper-parameter configuration with the lowest $ERDE_{50}$ value, $\lambda =\rho=1$ and $\sigma= 0.455$, we finally trained our model with the whole training set and performed the classification of the subjects from the test set.
Additionally, the classification of the test set was carried out applying two different classification policies, similarly to what was intuitively introduced in \autoref{sec:classification_process}:
the first one classified a subject as positive if the accumulated positive confidence value becomes greater than the negative one;
the second one, denoted by SS3$^\Delta$, was more comprehensive and classified a subject as positive when the first case was met, or when the change of the positive slope was, at least, four times greater than the negative one, i.e. the positive value increased at least 4 times faster\footnote{Those readers interested in the implementation details for this scenario, the classification algorithm is given in the next section.}. 
 The obtained results are shown in \autoref{tab:erisk-chunks} and are compared against each institution's best ERDE$_5$ and ERDE$_{50}$ among all the 30 submissions \footnote{The full list is available at \href{http://early.irlab.org/2017/task.html}{early.irlab.org/2017/task.html}.}.
It can be seen that SS3 obtained the best ERDE$_5$ (12.60\%) while SS3$^\Delta$ the best ERDE$_{50}$ (7.72\%).
Additionally, standard timeless measures were $F_1=0.52$, $\pi=0.44$ and $\rho=0.63$ for SS3 and $F_1=0.54$, $\pi=0.44$ and $\rho=0.69$ for SS3$^\Delta$. SS3$^\Delta$ had the 7th best $F_1$ value (0.54) out of the other 30 contributions and was quite above the average ($0.39$), which is not bad taking into account that hyper-parameters were selected with the aim of minimizing ERDE, not the $F_1$ measure.

\subsubsection{Scenario 2 - modified setting, incremental post-by-post classification} \label{subsec:exp-scenario2} 

As said earlier, each chunk contained 10\% of the subject's writing history, a value that for some subjects could be just a single post while for others hundreds or even thousands of them. Furthermore, the use of chunks assumes we know in advance all subject's posts, which is not the case in real life scenarios, in which posts are created over time. Therefore, in this new (more realistic) scenario, subjects were processed one writing (post) at the time (in a stream-like way) and not using chunks. 

Given that we do not have previous results available from other participants under this new scenario, for comparison, we had to perform experiments not only with SS3 but also with other standard classifiers Logistic Regression (LOGREG), Support Vector Machine (SVM), Multinomial Naive Bayes (MNB) and $K$-Nearest Neighbors ($K$-NN). For all these standard methods, the policy to classify a stream as positive (depressed) was the same as the most effective policy used in \citep{losada2016test}, that is, classify a subject as depressed when the classifier outputs a confidence value above 0.5.

As it will be discussed in the next section, when classifying a subject in a streaming-like way, the execution cost of each classifier for each subject is $O(n^2)$ with respect to the total number of subject's writings, $n$ ---except for MNB and SS3 which is $O(n)$.
In accordance to this, if we had used cross-fold validation to find the best parameters of each classifier to minimize the ERDE measure it would have taken too much time, more than one hour for every single fold and for every single possible combination of parameter values (i.e. weeks or even months in total). Therefore, parameters were selected with the aim of optimizing, as usual, the standard F$_1$ measure instead of ERDE.

Since the dataset was highly unbalanced we optimized the penalty parameter, $C$ $(C > 0)$, and the class weight parameter $w$  $(w \geq 1)$ for SVM and LOGREG; for MNB only the class weight $w$ was varied, while for $K$NN the $K$ parameter. 
As in \citep{losada2016test}, we set the majority class (non-depressed) weight to $1/(1 + w)$ and the minority class (depressed) weight to $w/(1 + w)$. Also, following standard practice, we applied a grid search on the tuning parameters, with exponentially growing sequences ($C = 2^{-10}, 2^{-4}, \cdots, 2^9$ and $w = 2^0, 2^1, \cdots, 2^9$) for SVM, LOGREG, and MNB and for the case of $K$-NN, $K$ took values sequentially from 1 to 20.

Model selection was also done by 4-fold cross-validation on the training data optimizing the F$_1$ measure with respect to the minority class. The parameter configuration with the highest F$_1$ for each classifier was the following: $C=16$ and $w=16$ for SVM (with L2 regularization); $C=16$ and $w=4$ for LOGREG (with L1 regularization); $w=1$ for MNB; $K=2$ for KNN; and $\lambda= 1.68$, $\rho=0.38$, and $\sigma=0.5$ for SS3.

We trained the classifiers using the optimized parameters with the whole training dataset and then, the incremental post-by-post classification was analyzed. Now, writings are processed sequentially, that is, early classification evaluation was carried out, as mentioned, one writing at a time. Additionally, we decided to compute the ERDE measure not only for $o=5$ and 50 but also for $o=10$, 30, 75 and 100 in order to have a wider view of how efficient classifiers are with respect to how early they classify subjects. The obtained results are shown in \autoref{tab:erisk-writings}.
There, in the last column, it is also included the time that each classifier required to classify all the subjects in the test set. As we can see, SS3 obtained the best $F_1$ and ERDE values for all the considered $o$ values except for ERDE$_5$.
On the other hand, SS3 has a precision($\pi$) value (0.42) relatively similar to the best one (0.47), obtained by SVM.
However, as we will discuss further in the next section, SS3 has a more efficient computation time in comparison with the remaining algorithms. For instance, it took SVM more than one hour (73.9 min) to complete the classification of the test set while it took SS3 a small fraction of it (roughly 5.3\%) to carry out the same task

\begin{figure}[t!]
    \begin{subfigure}{0.5\textwidth}
        \centering
        \includegraphics[width=90mm]{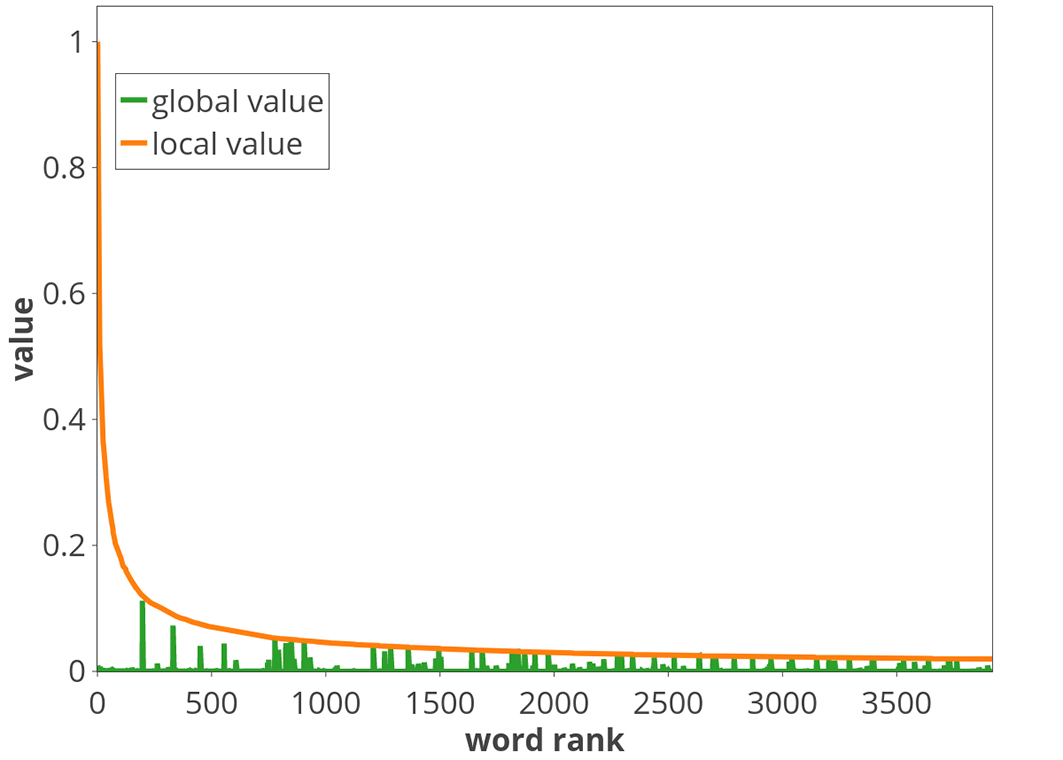}
    \end{subfigure}
\caption{
    \emph{global value} (green) in relation to the \emph{local value} (orange) for the ``depressed'' category. The abscissa represents individual words arranged in order of frequency.
    Note that the zone in which stop words are located (close to 0 in the abscissa) the  \emph{local value} is very high (since they are highly frequent words) but the \emph{global value} is almost 0, which is the desired behavior.
}
\label{fig:gv}
\end{figure}

\begin{table}[t!]
\centering
\caption{Results on the test set using all subject's history as a single document, i.e. timeless classification.}
\label{tab:erisk-fulldoc}
\begin{tabular}{l c c c}
\hline
& $F_1$ & $\pi$ & $\rho$ \\ \hline
SS3 & \textbf{0.61} & \textbf{0.63} & 0.60 \\
LOGREG & 0.59 & 0.56 & 0.63  \\
SVM & 0.55 & 0.5 & 0.62 \\
MNB & 0.39 & 0.25 & \textbf{0.96} \\
KNN & 0.54 & 0.5 & 0.58 \\ \hline

\hline
\end{tabular}
\end{table}

It is interesting to notice that we also performed classification of subjects on the test set using all subject's writings as if it were a single document (i.e. classical timeless classification); results are shown in \autoref{tab:erisk-fulldoc}.
SS3 obtained the highest values for $F_1$ (0.61) and Precision (0.63) measures, possibly due to the flexibility that is given by its three hyper-parameters to discover important and discriminative terms. These results provide strong evidence that SS3 also achieves competitive performance when is trained and tested to optimize standard (non-temporal) evaluation measures.  
Note that the best configuration of MNB obtained after the model selection stage, aiming at overcoming the unbalanced dataset problem, tends to classify all subjects as depressed, that is the reason MNB had a Recall($\rho$) close to 1 but a really poor precision (0.25).

\section{Analysis and Discussion}
\label{sec:analysisAndDiscussion}

\begin{figure*}[t!]
    \begin{subfigure}{0.5\textwidth}
        \centering
        \includegraphics[width=90mm]{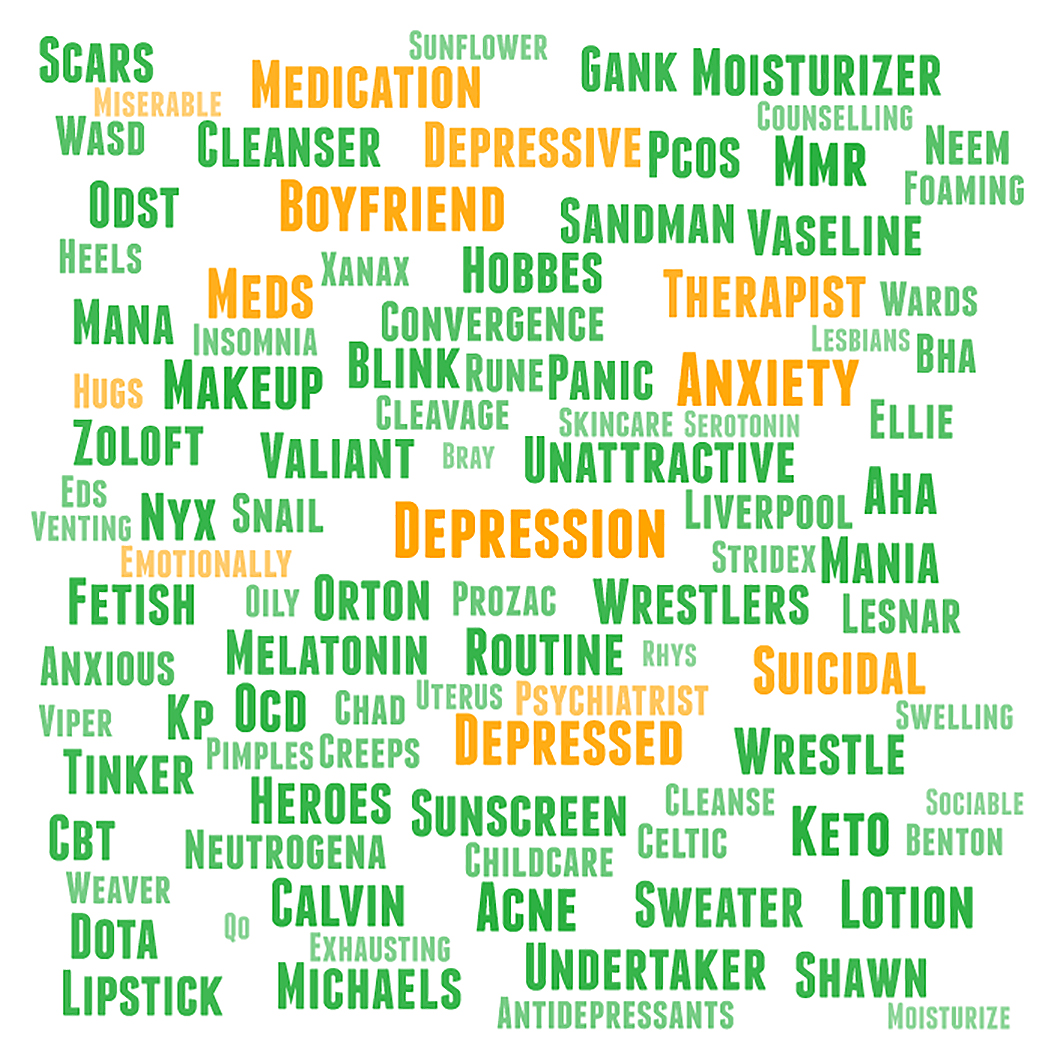}
        \caption{Sized by \emph{global value}}
        \label{fig:word_cloud_a}
    \end{subfigure}
    \begin{subfigure}{0.5\textwidth}
        \centering
        \includegraphics[width=90mm]{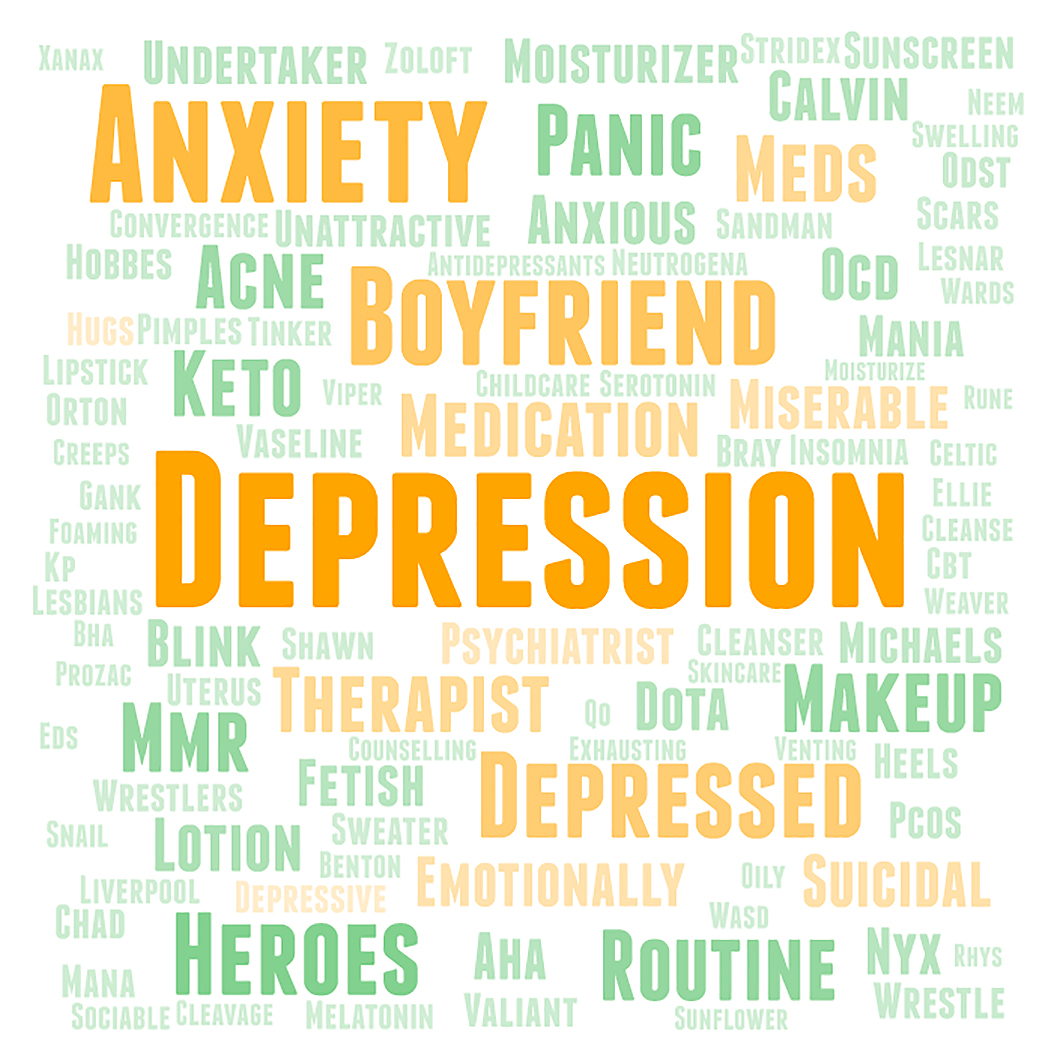}
        \caption{Sized by raw frequency}
        \label{fig:word_cloud_b}
    \end{subfigure}
\caption{
    Top-100 words selected by \emph{global value (GV)} from the model trained for the eRisk Pilot Task using chunks. The font size is related to (a) \emph{GV} and (b) row frequency. The green color indicates the words selected only by \emph{GV} whereas the orange color indicates the words also selected by the traditional Information Gain(\emph{IG}).
}
\label{fig:word_cloud}
\end{figure*}

From the experimental study of \autoref{sec:exp_and_results}, we can conclude that the proposed framework appears to show remarkable performance in incremental classification for early depression detection tasks. It obtained the best results for the time-aware error measures specifically designed to combine classifier’s accuracy and penalization in late classifications. In that context, it is important to notice that SS3 showed to be more effective than the others, more elaborated, approaches participating in the eRisk task, such as those based on Recurrent Neural Networks (like LSTM, GRU, etc.), graph-based models, ensembles of different classifiers, etc.

Regarding the support that SS3 provides for early classification we can say that, even though the rules we used are very simple, they are more effective than more elaborated and complex mechanisms used in the pilot task. For instance, some mechanisms to stop reading and classifying a subject included complex decision mechanisms based on specific rules for different chunks \citep{errecaldetemporal}. These rules take into account the decisions of different classifiers, the probability that each classifier assigned to its prediction, ``white lists'' containing the words with the highest information gain, and other sources of information.  Another approach that showed a good performance relied on hand-crafted rules specifically designed for this problem \citep{trotzek2017linguistic}, of the form: ``if output $\geq\alpha_n$ and number of writings $\geq n$, then classify as positive'', ``if output $\leq\beta_n$ and the number of writings $\geq n$, then classify as non-depressed'', etc.

As we can see, the two types of decision rules for early classification we used are quite simpler than those mechanisms and more importantly, they are problem-independent yet, interestingly, obtained better results in practice.
It is true that more elaborated methods that simultaneously learn the classification model and the policy to stop reading could have been used, such as in \citep{dulac2011text,yu2017learning}.
However, for the moment it is clear that this very simple approach is effective enough to outperform the remainder methods, leaving for future work the use of more elaborated approaches.

\

In order to get a better understanding of the rationale behind the good behavior of our framework, it is important to go into more details on the mechanisms used to weight words.
In \autoref{fig:gv} we can empirically corroborate that the \emph{global value} correctly captures the significance and discriminating power of words since, as it is well known, mid-frequency words in the distribution have both high significance and high discriminating power\footnote{As firstly hypothesized by \cite{luhn1958automatic}.}, and \emph{global values} for these mid-frequency words are the highest.

	This discriminating power of words can also be appreciated from a more qualitative point of view in the word-clouds of the top-100 selected words by \emph{global value} shown in \autoref{fig:word_cloud}.
	From this figure it is possible to observe that the most frequent terms, i.e. the biggest ones on (b), were also selected by $IG$ (orange colored), however, most of the terms selected only by $GV$ (green colored) are not so frequent, but highly discriminative.
	To highlight this point, note that \emph{GV} included very general words (depression, suicidal, psychiatrist, anxiety, etc.) but, unlike \emph{IG}, it also included many specific words.
	For instance: not only the word \emph{antidepressant} was included but also well-known antidepressants such as \emph{Prozac} and \emph{Zoloft}\footnote{Not included here, but also at rank 125 \emph{Lexapro}.};
	not only general terms related to medicine or disorders (such as \emph{medication, meds, insomnia, panic, mania}, etc.) but also more specific ones such as \emph{OCD} (Obsessive compulsive disorder), \emph{PCOS} (Polycystic ovary syndrome), \emph{EDS} (Ehlers-Danlos syndrome), \emph{CBT} (Cognitive behavioral therapy), \emph{serotonin}, \emph{melatonin}, \emph{Xanax}, \emph{KP} (Kaiser Permanente, a healthcare company), etc.;
	not only general words linked to diet, body or appearance (such as \emph{unattractive, skincare, makeup, acne}, etc.) but also \emph{pimples}, \emph{swelling}, \emph{Keto} (Ketogenic diet for depression), \emph{Stridex} (an American acne treatment and prevention medicine), \emph{AHA} (Alpha hydroxy acids), \emph{BHA} (Beta hydroxy acid), \emph{Moisturizer}, \emph{NYX} (a cosmetics company), \emph{Neutrogena} (an American brand of skin care, hair care and cosmetics), etc.
	It is also worth mentioning that this is a vital and very relevant aspect: if we value these specific words, as is usual, only by their local probability \footnote{Which is the case, for instance, with Multinomial Naive Bayes.} (or frequency), as shown in (b), they will always have almost ``no value'' since, naturally, their probability of occurrence is extremely small compared to more general words (and even worst against stopword-like terms). However, for instance, we intuitively know the phrase ``I'm taking antidepressants'' has almost the same value as ``I'm taking Prozac'' when it comes to deciding whether the subject is depressed or not.
	Fortunately, this is correctly captured by the \emph{global value}\footnote{Note that, unlike in (b), the size of ``Antidepressants'' and ``Prozac'' in (a), at the bottom and in the middle of it respectively, are quite similar and not so different from the size of ``Depression''.} since it was created to value terms, globally, according to how discriminative and relevant they are to each category.

\

Additionally, in order to better understand the good obtained results, another important aspect to analyze is how the early classification was actually carried out using the simplest policy to decide \emph{when} to positively classify subjects. In \autoref{fig:subjects_cases} are shown four subjects from the test set that illustrate four types of common classification behaviors we have detected:

\begin{figure*}[t!]
    \begin{subfigure}{0.5\textwidth}
        \centering
        \includegraphics[width=90mm]{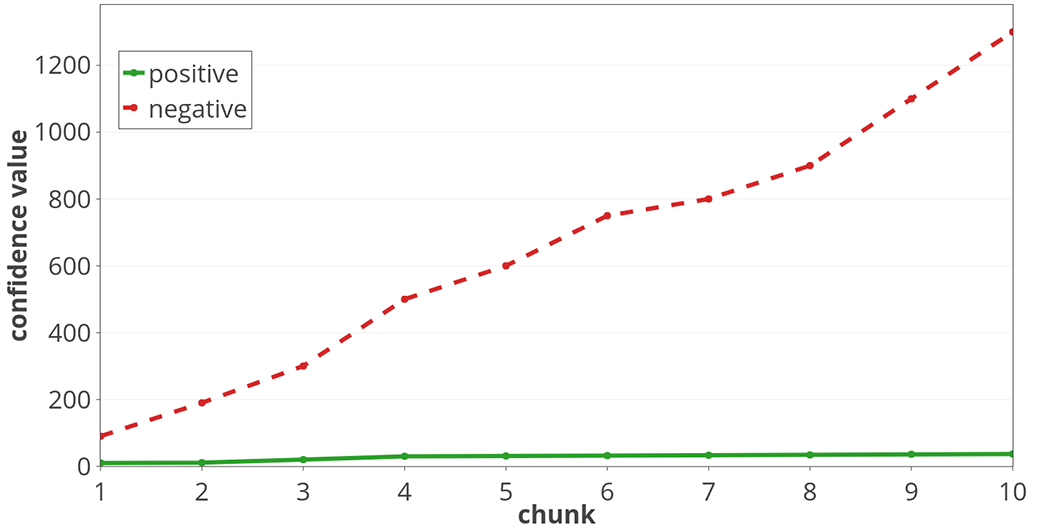}
        \caption{subject 265 (labeled as non-depressed)}
        \label{fig:subjects_gv_a}
    \end{subfigure}
    \begin{subfigure}{0.5\textwidth}
        \centering
        \includegraphics[width=90mm]{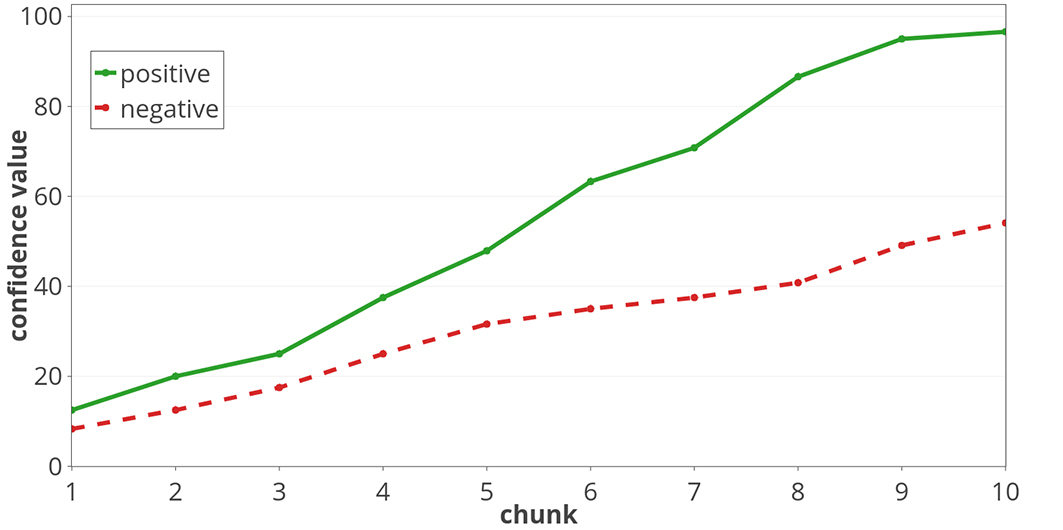}
        \caption{subject 9306 (labeled as depressed)}
        \label{fig:subjects_gv_b}
    \end{subfigure} \\
    \begin{subfigure}{0.5\textwidth}
        \centering
        \includegraphics[width=90mm]{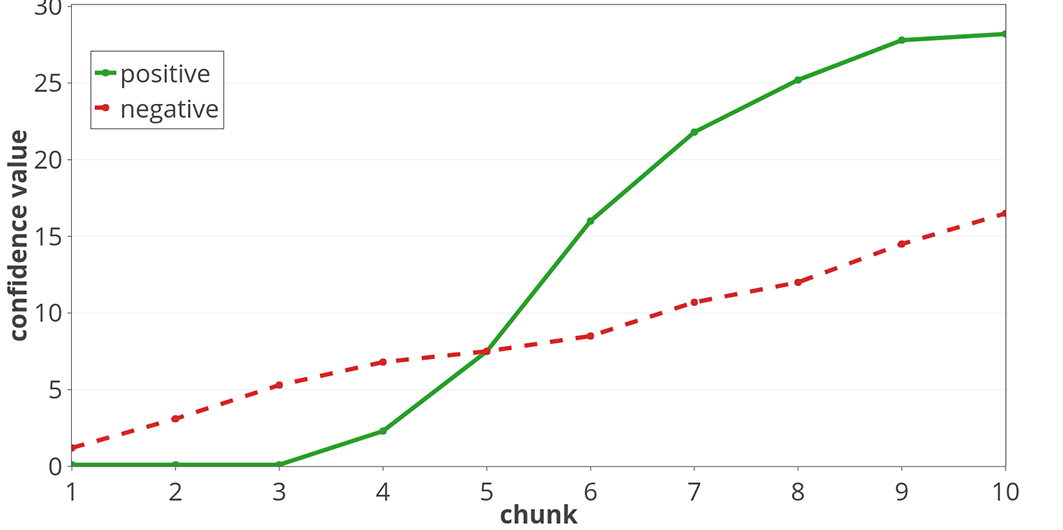}
        \caption{subject 9579 (labeled as depressed)}
        \label{fig:subjects_gv_c}
    \end{subfigure}
    \begin{subfigure}{0.5\textwidth}
        \centering
        \includegraphics[width=90mm]{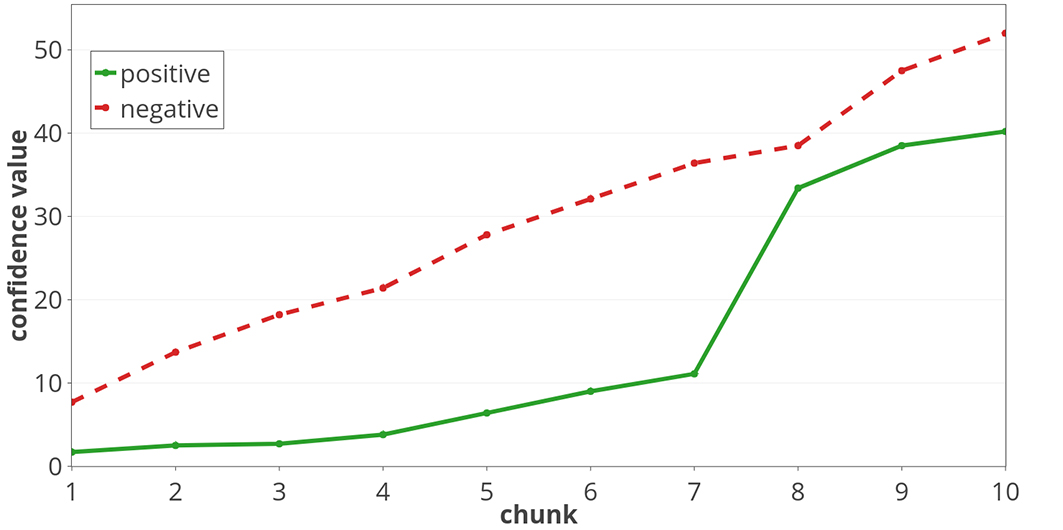}
        \caption{subject 1914 (labeled as depressed)}
        \label{fig:subjects_gv_d}
    \end{subfigure}
\caption{
Accumulated confidence values over time (chunk by chunk).
Four typical behaviors are shown, represented by these four subjects from the test set.
}
\label{fig:subjects_cases}
\end{figure*}

\begin{itemize}
\item[(a)]
from the first chunk on, the cumulative confidence value of one of the classes (negative in this case) stays above and always growing faster the other one. In this example, correctly, it was not possible to classify this subject as depressed after reading all its chunks.
\item[(b)]
similar to the previous case, the value of one class (positive) stays always on top of the other one, but this time they both grow at a similar pace. The subject was correctly classified as depressed.
\item[(c)]
the accumulated negative confidence value starts being greater than the positive one, but as more chunks are read (specifically starting after reading the 3rd chunk), the positive value starts and stays growing until it exceeds the other one. In this case, this subject is classified as depressed after reading the 6th chunk.
\item[(d)]
this example has a behavior similar to the previous one, however, the positive value, despite getting very close at chunk 8, never exceeds the negative one, which leads to the subject 1914 being misclassified as negative.
\end{itemize}

\begin{figure}[t!] 
    \centering
    \includegraphics[width=90mm]{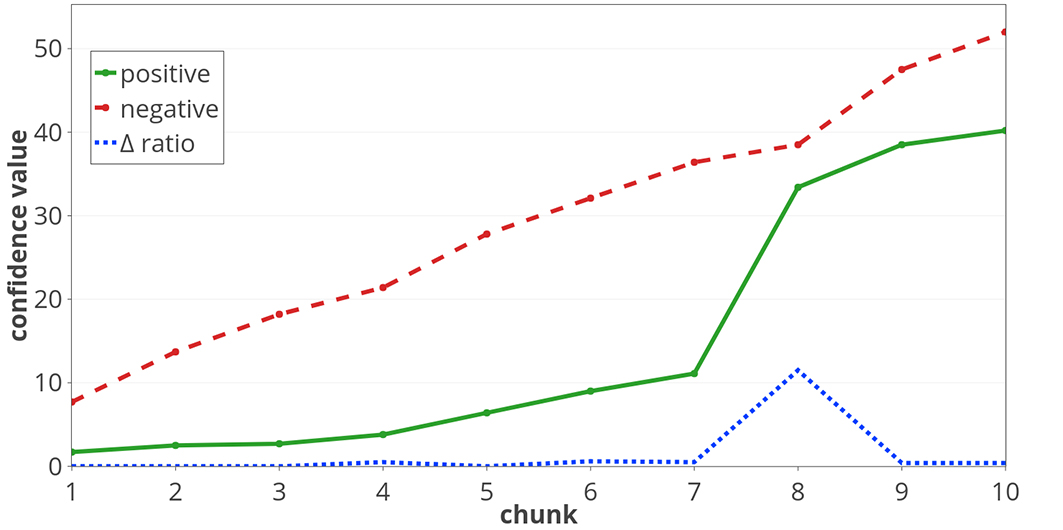}
    \caption{
    subject 1914 (labeled as depressed).
    The ratio between the positive and the negative slope change ($\Delta$) is shown in blue (dotted line). This ratio was used by the $\Delta$ policy.}
\label{fig:slope}
\end{figure}

With the aim of avoiding cases of misclassification like in (d), we decided to implement the second classifier, SS3$^\Delta$, whose policy also takes into account the changes in both slopes.
As it can be seen from \autoref{alg:classification_ss3slope} and as mentioned before, SS3$^\Delta$ additionally classifies a subject as positive if the positive slope changes, at least, four times faster than the other one.
In \autoref{fig:slope} is shown again the subject 1914, this time including information about the changes in the slopes.
Note that this subject was previously misclassified as not depressed because the accumulated positive value never exceeded the negative one, but by adding this new extra policy, this time it is correctly classified as positive after reading the 8th chunk\footnote{Note the peek in the blue dotted line pointing out that, at this point, the positive value has grown around 11 times faster than the negative one.}.

\begin{algorithm}[t!]
\small
\caption{\small
SS3$^\Delta$ classification algorithm.
Where \textproc{Classify-Chunk}($chunk$) is actually \textproc{Classify-At-Level}($chunk$, 4).
}
  \label{alg:classification_ss3slope}
  \begin{algorithmic}[h]
  \Statex
  \Function{Classify-Subject}{$chunks$}
      \State \textbf{input:} $chunks$, a subject's sequence of chunks
      \State \textbf{local variables:} $\overrightarrow{c}$, the subject \emph{confidence vector}
      \State \localvarsindent $\overrightarrow{\Delta c}$, a chunk \emph{confidence vector}
      \State $\overrightarrow{c} \gets (0, 0)$ $\rhd$ where (negative, positive)
      \For{\textbf{each} $chunk$ \textbf{in} $chunks$}
          \State $\overrightarrow{\Delta c} \gets $\Call{Classify-Chunk}{$chunk$}
          \State $\overrightarrow{c} \gets \overrightarrow{c} + \overrightarrow{\Delta c}$ 
          \If {$\Big(\frac{\overrightarrow{\Delta c}[1]}{\overrightarrow{\Delta c}[0]} > 4\Big)$ \Or $\Big(\overrightarrow{c}[1] > \overrightarrow{c}[0]\Big)$}
              \State \Return $positive$ \Comment{subject is depressed}
          \Else
              \State more evidence is needed
          \EndIf
      \EndFor
      \State \Return $negative$
  \EndFunction
  \end{algorithmic}
\end{algorithm}

\begin{figure*}[t!]
    \begin{subfigure}{0.5\textwidth}
        \centering
        \includegraphics[width=90mm]{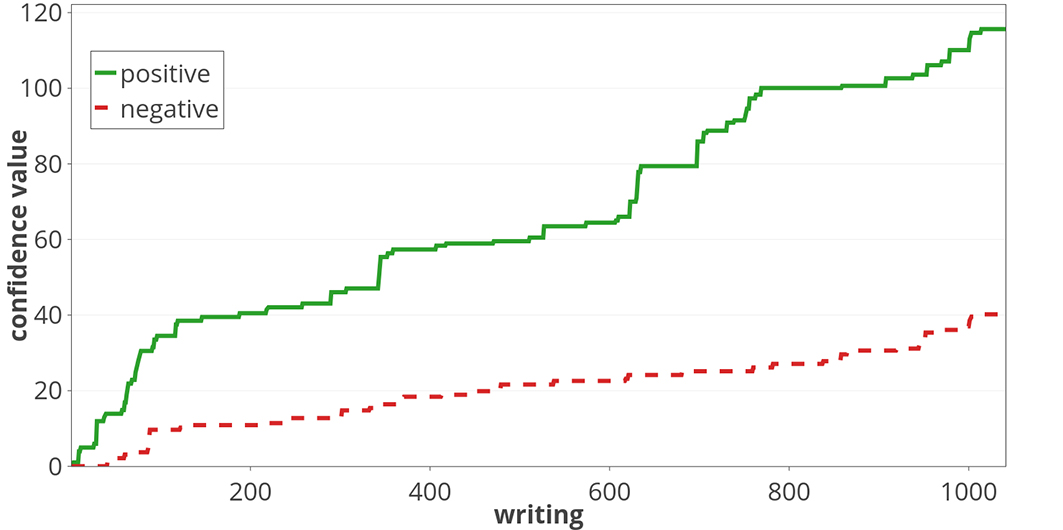}
        \caption{subject 834 (non-depressed).}
        \label{fig:subjects_error_a}
    \end{subfigure}
    \begin{subfigure}{0.5\textwidth}
        \centering
        \includegraphics[width=90mm]{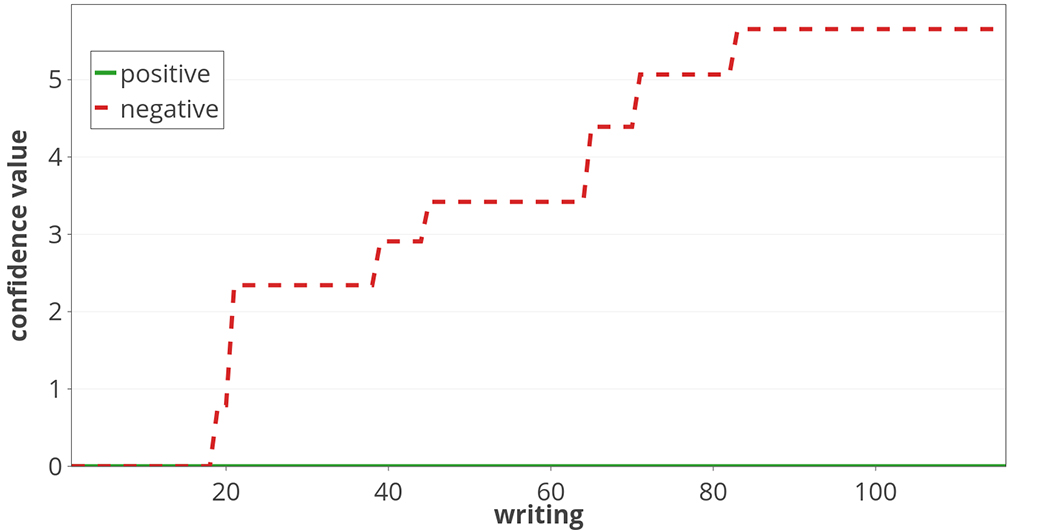}
        \caption{subject 1345 (depressed).}
        \label{fig:subjects_error_b}
    \end{subfigure} \\
    \begin{subfigure}{0.5\textwidth}
        \centering
        \includegraphics[width=90mm]{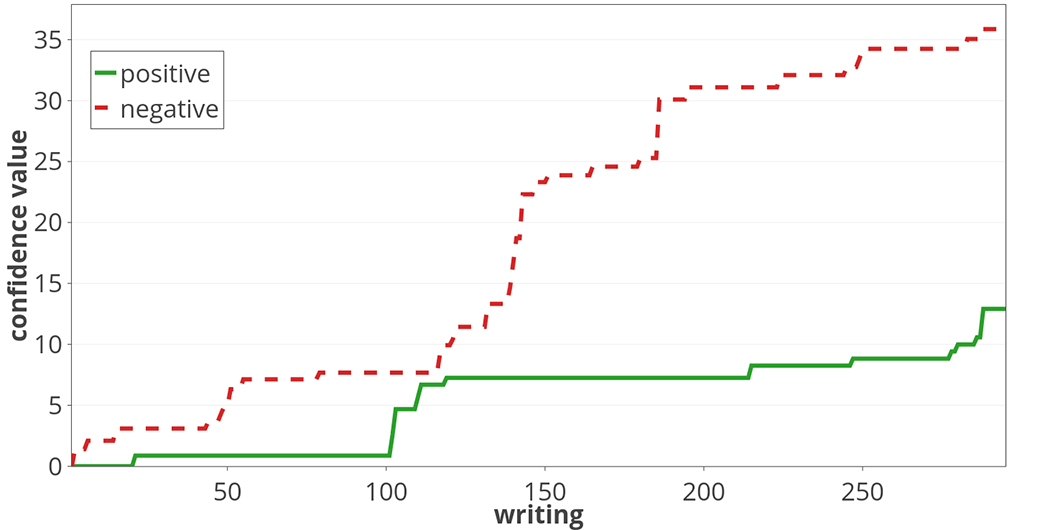}
        \caption{subject 2673  (depressed).}
        \label{fig:subjects_error_c}
    \end{subfigure}
    \begin{subfigure}{0.5\textwidth}
        \centering
        \includegraphics[width=90mm]{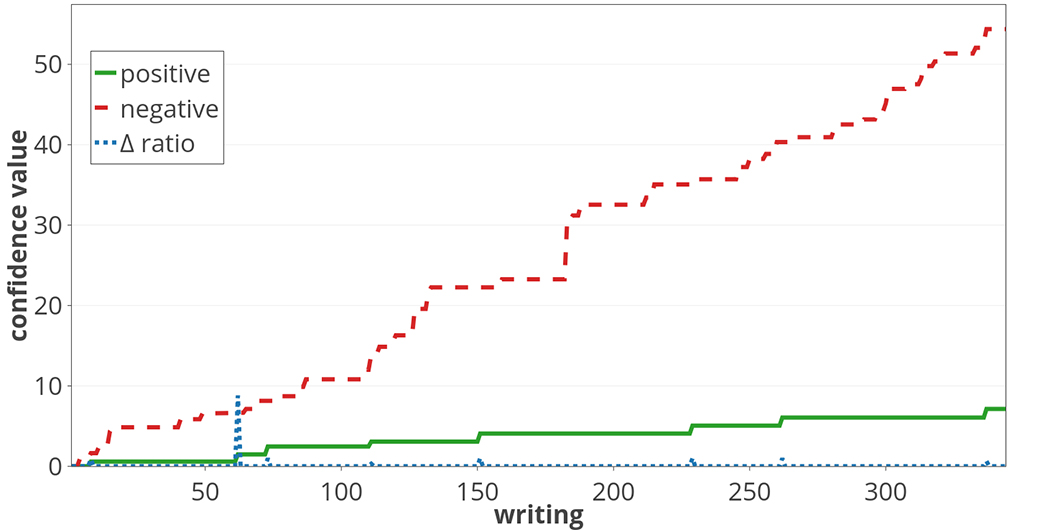}
        \caption{subject 748  (non-depressed).}
        \label{fig:subjects_error_d}
    \end{subfigure}
\caption{
Accumulated confidence values over time (writing by writing).
Four common error cases, represented by these four subjects.
}
\label{fig:subjects_error}
\end{figure*}

From the previous analysis, it is clear that useful information can be obtained from the study of those cases where our approach was not able to correctly predict a class. With this goal in mind, we also carried out an error analysis and identified four common error cases which could be divided into two groups: those that arise from bad labeling of the test set and those that arise from bad classifier performance. In \autoref{fig:subjects_error} we exemplify each case with one subject from the test set, described in more detail below:

\begin{itemize}
\item[(a)]
the subject is misclassified as positive since the positive accumulated exceeded the negative one. When we manually analyzed cases like these we often found out that the classifier was correctly accumulating positive evidence since the users were, in fact, apparently depressed.
\item[(b)]
in cases like this one, subjects were misclassified as negative, since SS3 did not accumulate any (or very little) positive evidence. Manually analyzing the writings, we often could not find any positive evidence either, since subjects were talking about topics not related to depression (sports, music, etc.).
\item[(c)]
there were cases like this subject, in which SS3 failed to predict ``depression'' due to the accumulated positive value not being able to exceed the negative one even although, in some cases, it was able to get very close. Note that the positive value gets really close to the negative one at around the 100th writing\footnote{Perhaps a finer tuning of hyper-parameters would overcome this problem.}.
\item[(d)]
this type of error occurred only due to the addition of the slope ratio policy.
In some cases, SS3 misclassified subjects as positive because, while it was true that the positive value changed at least 4 times more rapidly than the negative, the condition was mainly true only due to the negative change being very small.
For instance, if the change of the negative confidence value was 0.01, a really small positive change of at least 0.04 would be enough to trigger the ``classify as positive'' decision\footnote{Perhaps this could be fixed if we also request the positive or negative change to be, at least, bigger than a fixed constant (let us say 1) before applying the policy.}.
This problem can be detected in this subject by seeing the blue dotted peek at around the 60th writing, indicating that ``the positive slope changed around five times faster than the negative'' there, and therefore misclassifying the subject as positive. However, note that this positive change was in fact really small (less than 1). 
\end{itemize}

\begin{figure}[t!]
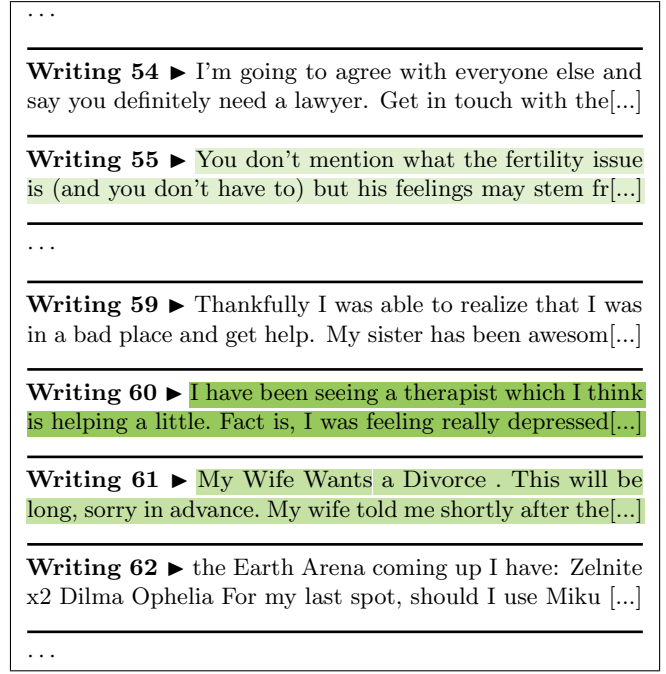
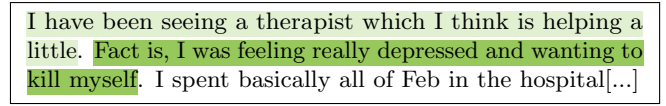
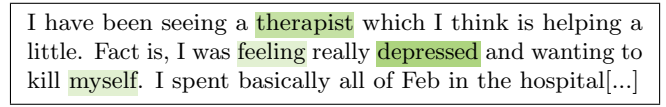

	\small
    \begin{subfigure}{90mm}
    	\centering
        \fbox{
        \begin{minipage}{0.9\textwidth}
        	\textbf{$\cdots$}\newline
        	\hlinew
            \textbf{Writing 54 $\blacktriangleright$} I'm going to agree with everyone else and say you definitely need a lawyer. Get in touch with the[...]\newline
			\hlinew
			\textbf{Writing 55 $\blacktriangleright$} \sethlcolor{dgreen_25}\hl{You don't mention what the fertility issue is (and you don't have to) but his feelings may stem fr[...]}\newline
            \hlinew
            \textbf{$\cdots$}\newline
            \hlinew
            \textbf{Writing 59 $\blacktriangleright$} Thankfully I was able to realize that I was in a bad place and get help. My sister has been awesom[...]\newline
            \hlinew
            \textbf{Writing 60 $\blacktriangleright$} \sethlcolor{dgreen_90}\hl{I have been seeing a therapist which I think is helping a little. Fact is, I was feeling really depressed[...]}\newline
            \hlinew
            \textbf{Writing 61 $\blacktriangleright$} \sethlcolor{dgreen_50}\hl{My Wife Wants a Divorce . This will be long, sorry in advance. My wife told me shortly after the[...]}\newline
			\hlinew
            \textbf{Writing 62 $\blacktriangleright$} the Earth Arena coming up I have: Zelnite x2 Dilma Ophelia For my last spot, should I use Miku [...]\newline
			\hlinew
            \textbf{$\cdots$}
        \end{minipage}
        }
        \caption{Subject 9579's history - writing level}
        \label{fig:subject9579_descriptive_a}
    \end{subfigure}
    \begin{subfigure}{90mm}
        \centering
        \fbox{
        \begin{minipage}{0.9\textwidth}
			\sethlcolor{dgreen_25}\hl{I have been seeing a therapist which I think is helping a little}. \sethlcolor{dgreen_90}\hl{Fact is, I was feeling really depressed and wanting to kill myself}. I spent basically all of Feb in the hospital[...]
       	\end{minipage}
        }
        \caption{Writing 60 - sentence level}
        \label{fig:subject9579_descriptive_b}
    \end{subfigure}
    \begin{subfigure}{90mm}
        \centering
        \fbox{
        \begin{minipage}{0.9\textwidth}
			I have been seeing a \sethlcolor{dgreen_50}\hl{therapist} which I think is helping a little. Fact is, I was \sethlcolor{dgreen_25}\hl{feeling} really \sethlcolor{dgreen_70}\hl{depressed} and wanting to kill \sethlcolor{dgreen_25}\hl{myself}. I spent basically all of Feb in the hospital[...]
       	\end{minipage}
        }
        \caption{Writing 60 - word level}
        \label{fig:subject9579_descriptive_c}
    \end{subfigure}
\caption{This figure shows how a visual description of the decision process could be given in this depression detection task. As we mentioned before, our framework allows us to analyze the reasons behind its classification decision, at different levels: (a) writings, (b) sentences and (c) words, etc. Each one of these blocks is painted proportionally to the real positive \emph{confidence values} we obtained after the experiments.}
\label{fig:subject9579_descriptive}
\end{figure}

Finally, we believe it is appropriate to highlight another of the highly desirable aspects of our Framework: its descriptive capacity.
As mentioned previously, most standard and state-of-the-art classifiers act as black boxes (i.e. classification process is not self-explainable) and therefore humans are not able to naturally interpret the reasons behind the classification.
However, this is a vital aspect, especially when the task involves sensitive or risky decisions in which, usually, people are involved. In \autoref{fig:subject9579_descriptive} is shown an example of a piece of what could be a visual description of the classification process for the subject 9579\footnote{Note that this is the same subject who was previously used in the example shown in \autoref{fig:subject9579_writings}, in \autoref{sec:classification_process}. The interested readers could see the relation between the green/positive curve there and the color intensity of each writing shown in \autoref{fig:subject9579_descriptive_a}.}.
In this example, we show in (a) a painted piece of the subject's writings history that the system users could use to identify which were the writings involved, and to what degree, in the decision making (classification) process. if the user wanted to further analyze, let us say, the writing 60 in more details, the same process could be applied at two different lower levels, as shown in (b) and (c) for sentences and words respectively.
It is worth mentioning that since this ``visual explanation'' process can be easily automated we have developed an online live demo, specially built for this purpose, available at \href{http://tworld.io/ss3}{http://tworld.io/ss3}. There, users can try out a version of SS3 trained using tweets for topic classification that, along with the classification result, gives a visual explanation.

\subsection{Computational Complexity}
\label{subsec:complexity}

As shown in \autoref{tab:erisk-writings}, SS3 is an efficient method in computation time.
This is due to the fact that, unlike most state-of-the-art classifiers, SS3 does not necessarily ``see'' the input as an atomic $n$-dimensional vector (i.e. a document vector) that must be computed entirely before making a prediction. In consequence, when working with a sequence of documents, for instance, SVM, LOGREG, and KNN must re-compute the input vector each time new content is added to the sequence.

Formally, if $n$ is the length of the sequence, when working with classifiers like SS3 or MNB, the cost of the early classification algorithm \emph{for every subject}, according to the number of processed documents, is equal to $n$ (since each document needs to be processed only once). On the other hand, for classifiers like SVM, LOGREG, KNN or (non-recurrent) Neural Networks, this cost is equal to $n\times(n+1)/2 = 1+2+...+n$ (since the first document needs to be processed $n$ times, the second $n-1$, the third $n-2$, and so on). Therefore, using the \emph{Big O Notation}, we have MNB and SS3 belonging to $O(n)$ whereas the other three classifiers belong to $O(n^2)$. Finally, it is worth mentioning that, as pointed out in the previous section, this cost affects not only the classification stage but it also severally affects previous stages such as hyper-parameter and model optimization since they need to classify the validation set several times (paying the cost every time).

It is worth noting that the difference in terms of space complexity is also very significant. For classifiers supporting incremental classification, like SS3 or MNB, only a small vector needs to be stored for each user. For instance, when using SS3 we only need to store the \emph{confidence vector}\footnote{In case of ADD, a 2-dimensional vector.} of every user and then simply update it as more content is created. However, when working with classifiers not supporting incremental classification, for every user we need to store either all her/his writings to build the document-term matrix or the already computed document-term matrix to update it as new content is added. Note that storing either all the documents or a $d\times t$ document-term matrix, where $d$ is the number of documents and $t$ the vocabulary size, takes up much more space than a small 2-dimensional vector.

Finally, since online social media platforms typically have thousands or millions of users, paying a quadratic cost to process each one while having to store either all the writings or a large $d\times t$ matrix for every user makes classifiers not supporting incremental classification not scalable.

\subsection{Implications and Clinical Considerations}
\label{sec:implications_clinical}

As stated in \citep{guntuku2017detecting}: ``Automatic detection methods may help to identify depressed or otherwise at-risk individuals through the large-scale passive monitoring of social media, and in the future may complement existing screening procedures''.
In that context, our proposal is a potential tool with which systems could be developed in the future for large-scale passive monitoring of social media to help to detect early traces of depression by analyzing users' linguistic patterns, for instance, filtering users and presenting possible candidates, along with rich and interactive visual information, for mental health professionals to manually analyze them. The ``large-scale passive monitoring'' aspect would be supported by the incremental\footnote{Only one small vector, the \emph{confidence vector}, needs to be stored for each user.} and highly parallelized nature of SS3 while the ``rich and interactive visual information'' one by its white-box nature.

It is clear that this work \emph{does not pursue a goal of autonomous diagnosis} but rather being a complementary tool to other well-established methods of mental health.
As a matter of fact, several ethical and legal questions about data ownership and protection, and how to effectively integrate this type of approaches into systems of care are still open research problems \citep{guntuku2017detecting}.

The dataset used in this task had the advantage of being publicly available and played an important role in determining how the use of language is related to the EDD problem. However, it exhibits some limitations from a methodological/clinical point of view. Beyond the potential ``noise'' introduced by the method to assess the ``depressed''/``non-depressed'' condition, it lacks some extra information that could be very valuable to the EDD problem.
For instance, in other datasets, such as the one used in \citep{de2013social} for the detection of depression in social media (Twitter in this case), in addition to the text of the interactions (tweets), it was also available other extremely valuable information for this type of pathology such as the scores obtained in different depression tests (CES-D and BDI), information about the user's network of contacts and interaction behavior (such as an insomnia index and posting patterns), among others.

It is clear that if we had had this additional information available, it would have been possible to obtain, among others, a more reliable assessment of depressive people, their severity levels of depression, and also to detect some mediating factors like environmental changes that could not be directly available in the users’ posts.
Besides, this information could also be used to train other models and integrate their predictions with the ones obtained only using textual information by, for instance, using some late-fusion ensemble approach.

Finally, although the clinical interpretability of the results was only addressed collaterally in our work, it is important to clarify some important points.
First of all, it was interesting to observe that most of the top-100 words relevant to the ``depression'' class, identified by our model, perfectly fit the usual themes identified in other, more clinical, studies on depression \citep{de2013social} such as ``symptoms'', ``disclosure'', ``treatment'' and, ``relationships-life''.
Interestingly, we also noticed what might be a new group of words, those linked to multiplayer online video games\footnote{As it can be seen in \autoref{fig:word_cloud} from words linked to the popular video game ``Dota'' like ``Dota'', ``MMR'', ``Wards'', ``Mana'', ``Rune'', ``Gank'', ``Heroes'' and ``Viper''.}, however, a reliable analysis of this requires a multidisciplinary work with mental health professionals that is out of the scope of the present work.
On the other hand, graphs of accumulated confidence values over time (chunk-by-chunk or writing-by-writing) shown in Figures 6, 7 and 8 are intended to show how lexical evidence (learned from the training data and given by $gv$) is accumulated over time, for each class, and how it is used to decide \emph{when} there is enough evidence to identify a subject as ``depressed''.
These figures should not be (mis)interpreted as trying to capture mood shifts or other typical behaviors in depressive people.

\section{Conclusions and Future Work}
\label{sec:conclusions}

In this article, we proposed SS3, a novel text classifier that can be used as a framework to build systems for early risk detection (ERD).
The SS3's design aims at dealing, in an integrated manner, with three key challenging aspects of ERD: incremental classification of sequential data, support for early classification and explainability.
In this context, we focused here on the two first aspects with a remarkable performance of SS3 (lowest $ERDE_o$ measure) in the experimental work with a very simple criterion for early classification.
SS3 showed better results than state-of-the-art methods with a more computationally efficient (O(n)) incremental classification process in two different scenarios, namely: incremental chunk-by-chunk and incremental post-by-post classification.
An additional interesting aspect was that it did not rely on (domain-specific) hand-crafted features neither on complex and difficult-to-understand mechanisms for early classification.
The SS3's virtue of being domain-independent contrasted with other effective algorithms for EDD which would require a costly process to adapt them to different problems.  Beyond that, we also showed with some intuitive examples, that the incremental/hierarchical nature of SS3 offers interesting support for explaining its rationale.

SS3 is a general and flexible framework that opens many research lines for future works. However, for the sake of clarity, we will focus here only on the more direct/evident ones. 

We believe that extending the predictive model by incorporating information related to non-linear aspects of human behavior, such as mood shifts, could help to capture when depression symptoms ``wax and wane''. This, for example, could help to detect when symptoms worsen as a means to prevent possible suicide or, if the subject is already diagnosed, to detect when applied therapy is not working. Having access to a dataset with this type of behavioral information would allow us in the future to integrate it into our EDD framework through, for example, a late-fusion ensemble approach.

Besides the limitations described in \autoref{sec:implications_clinical}, e.g. those caused by not using other information than text for classification, another limitation in the present work is that we used words as the basic building blocks (i.e. each writing was processed as a Bag of Words) on which our approach begins to process other higher level blocks (like sentences and paragraphs).
However, we could have been used different types of terms instead. For instance,  word $n$-grams could have helped us to detect important expressions (or collocations) that are not possible to identify as separate words, such as the ``kill myself'' in \autoref{fig:subject9579_descriptive_c}. Thus, in the future, we will measure how SS3 performs using other types of terms as well.

In the section ``Analysis and Discussion'' we could observe that  the \emph{global value} was a good estimator of word relevance for each category. We believe that this ability of  \emph{global value} to weight words could also play an important role as a feature selection method and, therefore, we will compare it against well-known feature selection approaches such as \emph{information gain} and \emph{chi-square} ($\chi^2$), among others.

Additionally, the framework flexibility and incremental nature allow SS3 to be extended in very different ways. Some possible alternatives could be the implementation of more elaborate \emph{summary operators}, $\oplus_j$, and more effective early stopping criteria.
Besides, with the aim of helping users to interpret more easily the reasons behind classification, for instance, for mental health professionals not familiar with the underlying computational aspects, we plan to continue working on better visualization tools.

Finally, the ``domain-independent'' characteristic of SS3 makes the framework amenable to be applied to other similar ERD tasks like anorexia, rumor or pedophile detection, among others. However, there is no impediment to use SS3 in other general author-profiling tasks (like gender, age or personality prediction) or even in standard text categorization tasks like, for instance, topic categorization.

\section*{\refname}
%
%








\bibliographystyle{model5-names}
\bibliography{manuscript}

%
\end{document}